\documentclass[12pt,preprint]{aastex}
\slugcomment{To appear in ApJ v655n2, February 1, 2007}

\shorttitle{DUST IN EARLY-TYPE GALAXIES}
\shortauthors{Sim\~{o}es Lopes et al.}

\newcommand{\hst}{{\it HST\,\,}}

\newcommand{\msun}{M$_\odot$}

\begin{document}

\title{A Strong Correlation between Circumnuclear Dust and Black Hole Accretion in Early-Type Galaxies} 
\author{Ramiro D. Sim\~{o}es Lopes, Thaisa Storchi-Bergmann, Maria de F\'{a}tima O. Saraiva}

\affil{Instituto de F\'{i}sica, Universidade Federal do Rio Grande do Sul, CP 15051, Av. Bento Gon\c{c}alves 9500, CEP 91501-970, Porto Alegre, RS, Brazil.}
\email{ramirosl@if.ufrgs.br}

\and

\author{Paul Martini}
\affil{Department of Astronomy, Ohio State University, 140 West 18th Avenue, Columbus, OH 43210}
\email{martini@astronomy.ohio-state.edu}

\begin{abstract}

We present a detailed investigation of the incidence of circumnuclear dust
structure in a large, well-matched sample of early-type galaxies with and
without Active Galactic Nuclei (AGN). All 34 early-type AGN hosts in our
sample have circumnuclear dust, while dust is only observed in 26\% (nine) of
a pair-matched sample of 34 early-type, inactive galaxies. This result
demonstrates a strong correlation between the presence of circumnuclear dust
and accretion onto the central, supermassive black hole in elliptical and
lenticular galaxies. This correlation is not present at later Hubble types,
where a sample of 31 active and 31 inactive galaxies all contain circumnuclear
dust. These archival, {\it Hubble Space Telescope} observations reveal a wide
range of mostly chaotic dust morphologies. Current estimates suggest the dust
settling or destruction time is on order of $10^8$ years and therefore the
presence of dust in $\sim 50$\% of early-type galaxies requires frequent
replenishment and similarly frequent fueling of their central, supermassive
black holes. The observed dust could be internally-produced (via stellar
winds) or externally-accreted, although there are observational challenges
for both of these scenarios. Our analysis also reveals that approximately a
third of the early-type galaxies without circumnuclear dust have nuclear
stellar disks. These nuclear stellar disks may provide a preferred kinematic
axis to externally-accreted material and this material may in turn form new
stars in these disks. The observed incidence of nuclear stellar disks and
circumnuclear dust suggests that episodic replenishment of nuclear stellar
disks occurs and is approximately concurrent with the fueling of the central
AGN.

\end{abstract}


\section{Introduction}

It is now widely accepted that most, if not all, galaxy bulges host a 
supermassive black hole (SMBH) at their center 
\citep{ferrarese00,gebhardt00,tremaine02}. However, only a fraction of these 
galaxies show evidence of significant accretion onto these central black holes 
in the form of Active Galactic Nuclei (AGN). 
The absence of significant accretion in many galaxies may be due to the lack 
of fuel (most likely in the form of cold interstellar gas), the lack of a 
fueling mechanism to remove angular momentum and drive this matter toward the 
center, or both. 
The most direct probe of the presence of fuel is the direct observation of 
cold interstellar gas, such as in the form of dust clouds in the central 
regions of galaxies. 
Cold dust is also dissipative and therefore sensitive to perturbations that 
may drive fuel toward the center of the galaxy. 
Thus the study of the dust distribution and its dynamical properties may 
provide valuable information about the powering, transport mechanisms, and 
timescales of nuclear activity.
Identification of the origin of AGN fuel and the nature of the triggering 
mechanism(s) are two of the main unsolved questions in AGN and black hole 
research \citep[e.g.,][]{martini04}. 

A large fraction of all galaxies, including those with early-type morphology, 
have long been known to possess some form of dust structure 
\citep[e.g.,][]{kormendy89}. 
The best spatial resolution of the centers of nearby galaxies can be obtained 
with the {\it Hubble Space Telescope} (\hst), which reveals the presence of 
cold interstellar clouds within parsecs of many nearby galactic nuclei. One of 
the first large \hst\ studies of early-type galaxies by \citet{vandokkum95} 
inferred that 78\% of a sample of 64 galaxies contain circumnuclear dust.  
These authors also noted that the frequency of dust detections is higher in 
radio-detected galaxies (72\%) than in galaxies without radio emission (33\%), 
which suggested a correlation with nuclear activity. 
More recent \hst\ studies have also found dust is common in the centers of 
early-type galaxies and a correlation between the presence of dust and 
line-emitting gas and/or AGN \citep{ravindranath01,tran01,lauer05}. 
For later Hubble types, \citet{malkan98} also showed that dust structures were 
present in most \hst\ WFPC2 images of a sample of 256 Seyfert and starburst 
galaxies; only a small fraction of the Seyfert galaxies did not appear to 
contain dust. Multicolor $V-$ and $H-$band observations of additional Seyferts 
further showed that circumnuclear dust was always found in the circumnuclear 
region of the predominantly spiral galaxy hosts of nearby Seyferts 2s 
\citep{regan99,martini99} and Seyfert 1s \citep{pogge02}.  

A more robust way to address the relation between the presence of circumnuclear 
gas and/or dust and nuclear activity is to look for differences between AGN 
host galaxies and a control sample of inactive counterparts. 
In a recent paper, \citet{xilouris02} searched for dust structures 
in \hst\ images of a sample of 23 Seyfert and 35 inactive galaxies by fitting 
and subtracting elliptical isophotes. They found that all active galaxies 
possess some form of nuclear dust structure and, for their subsample of 
early-type galaxies, the Seyferts exhibit more nuclear structure than the 
inactive galaxies, although the opposite seems to be true for the later types. 
In a more recent paper, \citet{martini03} investigate the incidence of 
dust in a well-matched sample of 28 active and 28 inactive galaxies and found 
that all of the AGN possess dust structure, while seven of the inactive 
galaxies did not. 

Studies of the centers of early-type galaxies have also revealed that many 
contain disky isophotes, and in some cases these isophotes are most likely due 
to embedded stellar disks 
\citep[e.g.,][]{capaccioli87,bender88,kormendy89,scorza95,seifert96}. 
Observations with \hst\ have identified many stellar disks within the 
central kiloparsec (hereafter nuclear stellar disks) that are distinct from 
the outer disks present in lenticular galaxies 
\citep{vandenbosch98,scorza98,morelli04,krajnovic04}. 
With a large sample of 67 early-type galaxies, 
\citet{rest01} found nuclear stellar disks in as many as 51\% of their sample, 
although many were misaligned and likely nuclear stellar bars. Given that 
disks are easiest to detect when nearly edge on, the observed fraction is 
consistent with the presence of nuclear stellar disks in all early-type 
galaxies \citep{rix90}. 

In the present paper we address the relation between circumnuclear dust 
structures and the nuclear activity in galaxies, with a particular emphasis 
on early-type galaxies. Unlike most previous work, we use a well-defined and 
homogeneous input sample -- the Palomar sample of \citet{ho95} -- and have 
carefully drawn a pair-matched sample of 34 active and 34 inactive early-type 
galaxies (ellipticals and lenticulars), matched by the properties of their host 
galaxies, in order to obtain a robust measurement of the frequency of dust 
structures in active and inactive early-type galaxies. 
We have also identified and analyzed a pair-matched sample of 31 active and 31 
inactive late-type galaxies. This paper is organized as follows: In Section 
\ref{sec-sample} we present the sample selection and the pair-matching 
technique; in Section \ref{sec-data} we present data reduction and analysis 
procedures; in Sections \ref{sec-results} and \ref{sec-discussion} we present 
our results and discussion, respectively, and finally in Section 
\ref{sec-conclusions} we present our conclusions.

\section{Sample\label{sec-sample}}

A key issue in the comparison of active and inactive galaxies is the 
selection of 
a well-defined control sample known to be inactive. This is difficult 
because identification of a low-luminosity AGN often requires less sensitive 
observations than confirmation that a low-luminosity AGN is not present.
In order to avoid this problem, we have drawn our sample from the large, uniform, 
and sensitive Palomar Survey of \citet{ho95}. 
The Palomar Survey contains spectra of the closest 486 
bright galaxies in a set region of the sky and all of the nuclear spectra 
are classified as: absorption-line nuclei, HII, LINERs, transition objects, 
and Seyfert galaxies \citep{ho97a}. The Palomar Survey is considered to be 
the most complete and homogeneous representation of the nearby universe. 
Our first selection of the active galaxy sample comprised all Palomar 
Seyfert and LINER galaxies with available broad-band \hst\ WFPC2 images in 
the optical spectral region, excluding transition objects and those with 
uncertain classification. The sample of inactive galaxies comprised all 
Palomar galaxies classified as absorption-line and HII nuclei with available 
broad-band WFPC2 images in the optical spectral region.

We then carefully selected a well-matched control sample from this sample 
of inactive galaxies through use of a pair-matching algorithm. This approach 
is similar to that adopted by \citet{martini03}, who argue that the best way 
to match control samples is to identify a control galaxy pair for each active 
galaxy with similar values of all properties that may affect the identification 
of dust structure (e.g., morphology, distance, luminosity, inclination).  
More traditional and commonly employed techniques, in contrast, just match 
the mean or median value of each property between the target and control 
samples, or the distribution of each property individually, and may be 
susceptible to different correlations between various properties in the two 
samples. 

\begin{figure}
\plotone{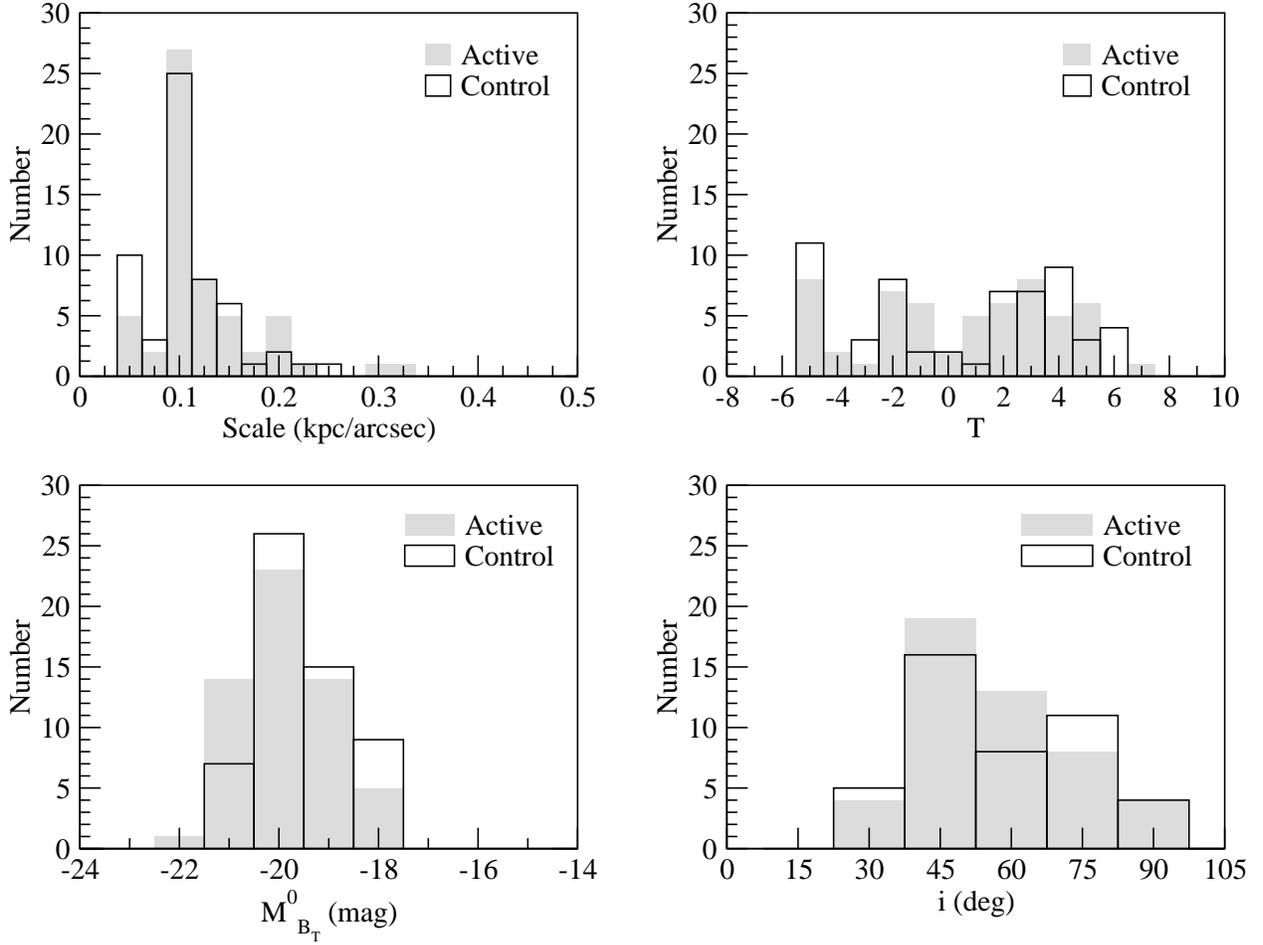}
\caption{Distributions of plate scales, Hubble types, absolute B magnitudes, and inclinations
for all active and control galaxies of the matched sample.\label{fig-hmatchsample}}
\end{figure}

\begin{figure}
\plotone{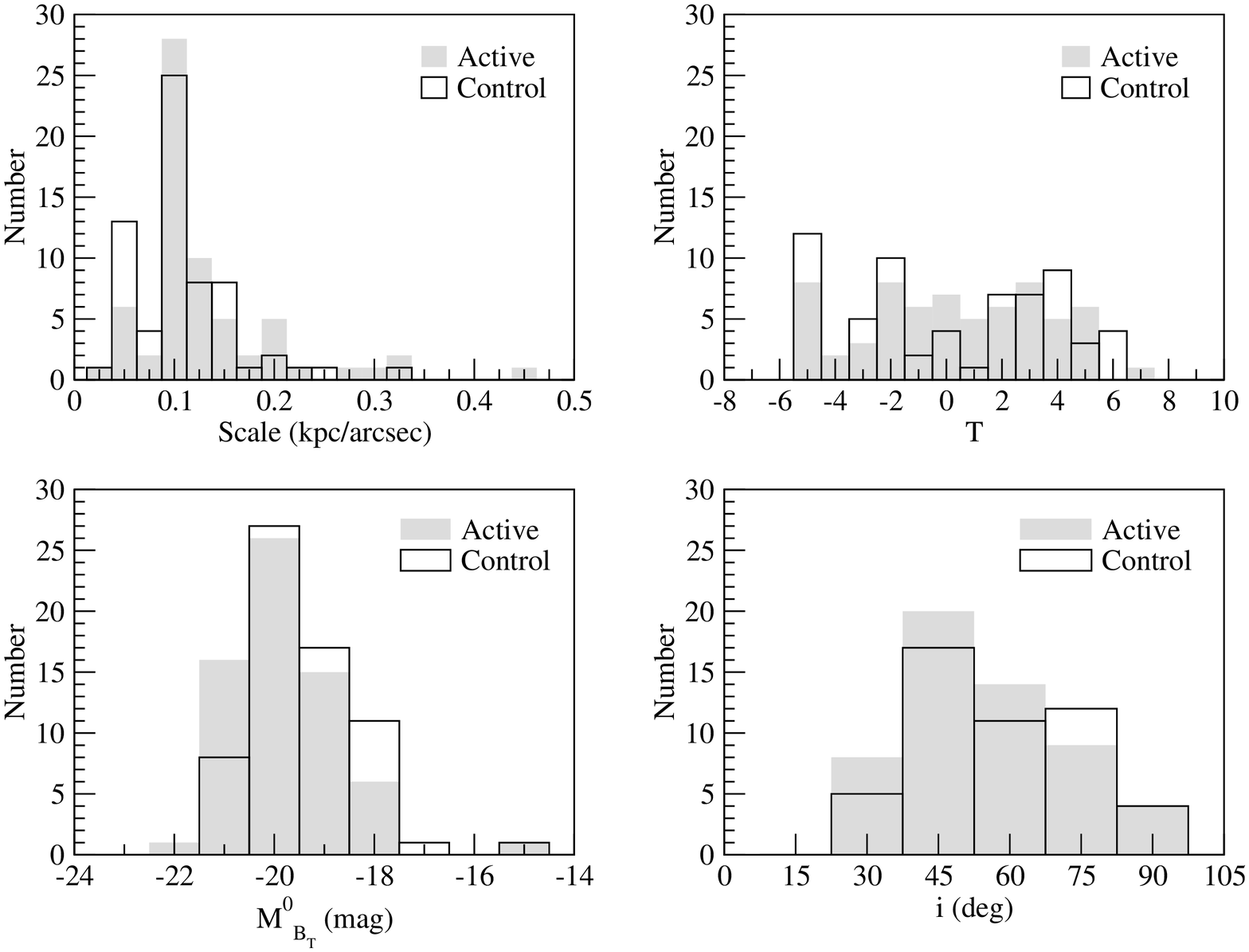}
\caption{Distributions of plate scales, Hubble types, absolute B magnitudes, and inclinations
for all active and control galaxies of the extended sample.\label{fig-hextsample}}
\end{figure}

In order to make a robust comparison between active and inactive galaxies we 
have identified an inactive pair for each active galaxy with the following 
criteria: 
maximum difference between absolute magnitudes of 1\,mag; maximum difference 
between morphological $T$ types of 1; maximum difference between galaxy 
inclinations of 15$\arcdeg$; maximum difference in plate scale of 50\%. To 
accomplish this we developed an algorithm that finds the largest possible 
number of pairs between the active and inactive galaxies given the above 
criteria. This yielded 26 early-type and 31 late-type pairs. These 57 pairs 
comprises our main sample and we shall hereafter refer to it as the ``matched 
sample''. In order to increase the early-type sample we then reran the pairing 
algorithm with the remaining early-type galaxies after relaxing the matching 
constraints to: maximum difference between absolute magnitudes of 3\,mag; 
maximum difference between $T$ types of 3; maximum difference between galaxy 
inclinations of 35$\arcdeg$; maximum difference in plate scale of 80\%. 
This resulted in 8 additional early-type pairs for a total of 
34 matched pairs of early-type galaxies. This sample, combined with our 
31 late-type pairs, we shall refer to as the ``extended sample.'' Histograms 
of the Hubble types, distances, inclinations and absolute $B$ magnitudes for 
the matched and extended sample are presented in Figures~\ref{fig-hmatchsample} and
\ref{fig-hextsample} respectively.

These figures show there are similar distributions for each property in the 
active and inactive galaxies. Although this is expected for the matched sample, 
it also remains true for the extended sample, and demonstrates that the 
additional 8 early-type galaxy pairs in the extended sample are reasonably 
well-suited for comparison of the properties of active and inactive early-type 
galaxies. 
Relevant information for the sample galaxies is presented in 
Table~\ref{tab-active} for the active galaxies and Table~\ref{tab-control} 
for the inactive galaxies. Columns 5-9 of each of these tables list the 
galaxy $T$ type, distance, absolute magnitude, inclination, and activity type 
from \citet{ho97a}. 

\section{Data Reduction and Analysis\label{sec-data}}

These data comprise \hst\ WFPC2 images obtained both with the planetary and 
wide field cameras. Relevant data on the images, notably the camera, filter, 
and exposure times, are listed in columns 2-4 of Tables~\ref{tab-active} and 
\ref{tab-control}. All of the images were initially reduced by the \hst\ 
archive's on-the-fly reprocessing pipeline. The only additional processing was 
cosmic-ray removal with the appropriate IRAF tasks. Specifically, when two or 
more images were available we used the CRREJ task in the STSDAS package, 
otherwise we used the COSMICRAY task in the NOAO package.

To identify circumnuclear dust structure in these images we used the 
structure map technique proposed by \citet{pogge02}. Structure maps 
enhance structure as fine as the scale of the point spread function in an image 
and are well suited to the identification of narrow dust lanes and emission-line
gas in nearby galaxies. The technique is based on the Richardson-Lucy (R-L) 
image restoration algorithm, which uses the PSF to identify structure to 
enhance as part of the deconvolution process. Mathematically, a structure map 
$S$ is defined as: 
\begin{equation}
	S = \left[ \frac{I}{I \otimes P} \right] P^t
\end{equation}
where $I$ is the original image, $P$ is the instrument PSF \citep[constructed 
with the TinyTim software of][]{krist04}, $P^t$ is the transpose of the PSF and 
$\otimes$ is the convolution operator \citep[see][]{pogge02}.

The structure map technique recovers similar information to the more 
traditional methods of contrast enhancement, such as fitting and subtracting 
elliptical isophotes and construction of color maps, but offers several 
advantages. For example, the fitting and subtraction of ellipses can lead to 
artificial features in the presence of strong brightness discontinuities or 
isophotal twists in the images. In the case of color maps, there are simply not 
as many galaxies with observations in two filters as in one. 
Possible mismatches between the PSFs in the two bands could also introduce 
artificial features in color maps. These problems are all avoided with the 
use of structure maps, although some spurious features can still appear. 
For example, overexposed pixels can produce dark rings that mimic absorption 
features, while dead pixels can appear as false bright spots in the images. 
Nevertheless, such cases were easily identified and did not affect our 
analysis. After constructing structure maps, we then visually inspected each 
galaxy to determine if dust structures were present and measured 
the spatial extent of any dust. We also noted the presence of emission 
features that seem to be nuclear stellar disks (see discussion below) and 
measured the projected radial extent of these features. 

\section{Results\label{sec-results}}

Structure maps for all our galaxies are shown in Figures~\ref{fig-stmap1} to 
\ref{fig-stmap7}. Figure~\ref{fig-stmap1} present 
structure maps of each early-type ($T \leq 0$), AGN host next to its 
inactive, control galaxy for the matched sample. Each galaxy pair has the 
active galaxy on the left and the inactive galaxy on the right. These two 
figures illustrate the most striking result of this paper, namely that all 
of the early-type active galaxies possess circumnuclear dust while only 
27\% (7 of 26) of the early-type inactive galaxies possess circumnuclear dust. 
Figure~\ref{fig-stmap4} demonstrates that this result also holds when we 
include the extended sample, specifically all of the early-type active galaxies have 
circumnuclear dust and only 26\% (9 of 34) of the inactive galaxies in the 
extended sample have circumnuclear dust. 
This result also holds if we exclude lenticulars and only consider 
ellipticals ($T = -5$). Eight of eight (10 of 10) active ellipticals in our 
matched (extended) sample have dust, while only one of eight (two of 15) 
inactive ellipticals in our matched (extended) sample have circumnuclear dust. 

While previous investigators have noted an excess of circumnuclear dust 
in early-type, active galaxies, compared to early-type, inactive galaxies, 
ours is the first study to compare the frequency of circumnuclear dust 
between well-matched samples. In addition, both our active and inactive, 
control samples have sensitive and uniform spectroscopy from the Palomar 
survey. These data are important to confirm both the presence of activity and 
its absence in the control sample. Some of the earliest evidence that 
early-type, active galaxies may more commonly possess dust came from 
studies of radio galaxies \citep{sadler85,marston88,jaffe94}. 
\citet{vandokkum95} presented the first study of the incidence of dust in 
active and inactive early-type galaxies with \hst\ and found dust in 72\% of 
the radio galaxies, while only in 33\% 
of the galaxies without radio emission. Other investigators \citep{kleijn99,
ravindranath01} have similarly found circumnuclear dust in most 
active galaxies, while rarely found dust in inactive, early-type galaxies. 
The \citet{ravindranath01} sample is the most comparable to our own because 
that sample was also selected from the Palomar survey, although their sample 
is somewhat smaller (33 galaxies total), did not match active and inactive 
hosts, and not all galaxies possessed visible-wavelength \hst\ images. 
We attribute our 100\% detection rate of dust in early-type AGN hosts to 
the excellent sensitivity of the Palomar Survey to low-luminosity AGN. 

The late-type ($T > 0$) galaxies are shown in Figure~\ref{fig-stmap5}.
All of these galaxies display evidence for circumnuclear 
dust regardless of the presence or absence of detected nuclear activity. 
Previous studies of dust in galaxies with similar Hubble types had shown 
that circumnuclear dust is common in active, late-type galaxies 
\citep[e.g.,][]{malkan98,regan99,martini99,pogge02}. Circumnuclear dust 
has also long been known to be quite common in late-type, inactive galaxies. 
With a similarly matched sample of 28 active and 28 inactive galaxies, 
\citet{martini03} found that all active galaxies had circumnuclear dust, 
although 25\% (7 of 28) of the inactive galaxies did not show evidence for 
circumnuclear dust. Only three of the seven inactive galaxies in their 
sample have later Hubble type than S0, therefore their measured dust frequency 
is consistent with the presence of dust in all 31 of our late-type, inactive 
sample.
The frequency of circumnuclear dust structure as a function of Hubble type for 
the matched and extended samples are shown in Figure~\ref{fig-hstructure}. 

Most of the dust features in these early-type galaxies are dominated by dust 
lanes and in many cases the lanes trace some spiral structure that indicates 
the presence of a dusty disk. In total, at least half of the active early-type 
galaxies exhibit some evidence for dust disks and 25\% of these have tightly 
wound spirals in regular disks. These later dust structures are interesting 
because tightly wound dust structures are generally observed in the most 
axisymmetric (i.e., unbarred) late-type galaxies and suggest the presence of 
long-lived structures \citep{peeples06}. 
The presence of any dust structures in each galaxy is indicated in column~11 
of Tables \ref{tab-active} and \ref{tab-control}, where we report the projected 
radial extent of each dust feature in kiloparsecs. This radius was estimated 
by visual inspection of the digital images. A $\times$ sign means that the 
radius of the structure is not clear, and may extend beyond the borders of 
the image. For the late-type galaxies we have not measured the radius of the 
dust structure as in most cases it does extend beyond the borders of the 
images. The dust disks and spirals have a mean size of 0.43$\pm$0.37\,kpc
and 0.69$\pm$0.63\,kpc respectively for the early-type active galaxies.
These structures therefore appear to be confined to the circumnuclear region 
or central kiloparsec, although these observations are less sensitive to 
dust on larger scales due to a decrease in signal-to-noise ratio. The 
position angles of the dusty disks in the early-types, if apparent, are 
generally aligned with the position angle of the large-scale isophotes. 
Some exceptions include orthogonal dust disks in NGC~2787, NGC~4111, 
and NGC~4143. 

In addition to circumnuclear dust, our structure maps also reveal the presence 
of nuclear stellar disks in many galaxies. We have found disks 
in at least 13 of the 34 (38\%) early-type inactive galaxies, while we
found nuclear stellar disks in only one of the 34 (3\%) early-type active 
galaxies. Many of these nuclear stellar disks were identified in previous 
work, such as NGC\,4570, NGC\,4621 and NGC\,5308 by \citet{krajnovic04}, which 
employed elliptical isophote fits to identify these disks. As we have 
identified a similar fraction of stellar disks, structure maps may be as 
sensitive to nuclear disks as isophote fitting. This is somewhat surprising, 
as structure maps are best suited to the detection of features on the scale of 
the point spread function. The appearance of the stellar disks in the structure 
maps may therefore indicate that they are nearly edge on, as is the case for 
elliptical isophote fits. Consistent with this interpretation, nearly all of 
the nuclear stellar disks in our sample have very large axis ratios. The 
detectability of stellar disks in early-type galaxies was modeled in detail by 
\citet{rix90}, who found that the observed fraction is consistent with the 
presence of a nuclear stellar disk in every early-type galaxy. 

The apparent nuclear stellar disks in these galaxies could also be 
nuclear stellar bars \citep[e.g.,][]{erwin02}. To determine if this 
is the case we measured the position angle of each of the nuclear stellar 
disks and compared it with the position angle of the galaxy isophotes on 
large scales. In all cases the nuclear stellar disk was aligned with the 
position angle on larger scales, which suggests that they are not nuclear 
bars. Surprisingly, the position angles of the nuclear stellar disks 
in five out of five elliptical galaxies 
also agree with the position angle of the isophotes at large scales 
(NGC 821, NGC 3610, NGC 3377, NGC 4621, and NGC 7619). 
While some of these ellipticals are known to be disky (e.g., NGC 3377, 
4621), ellipticals are known to be triaxial and do not all rotate about their 
apparent minor axis \citep{ryden92}. We return to this point in 
Section~\ref{sec:ndisks}. 

In contrast to the early-type, inactive sample, we do not detect many nuclear 
stellar disks in the early-type active sample or in either of the late-type 
subsamples. The absence of nuclear disks in late-type galaxies has already 
been reported in the work of \citet{pizzella02}, who studied \hst\ WFPC2 F606W 
images of a sample of 38 spiral galaxies and  did not find nuclear disks in 
barred galaxies or galaxies of Hubble type later than Sb. These differences 
from our early-type, inactive sample are most likely due to the substantial 
circumnuclear dust in the early-type active sample and the late-type galaxies. 
In almost all cases, the dust (and emission-line regions) present in these 
galaxies make it difficult to identify nuclear stellar disks, particularly 
if the dust and stellar disks have the same position angle. Some notable 
exceptions are NGC~4111, which has a prominent stellar disk almost 
perpendicular to a smaller dust disk, and NGC~4026, which has a nuclear 
stellar disk and weak dust features. If nuclear stellar disks are also present 
in the dusty, active early-types, the fraction may be higher in a study of 
near-infrared observations of early-type galaxies, although as 
\citet{ravindranath01} note the nuclear stellar disk fraction may also 
depend on the central surface brightness profile. 

A histogram of the presence of stellar disks as a function of Hubble type in 
our matched and extended samples is shown in Figure~\ref{fig-hdisk}. 
The presence of nuclear stellar disks in our sample is identified in column 12 
of Tables~\ref{tab-active} and \ref{tab-control} and provides the radial extent 
of the structure in kiloparsecs. In Figure~\ref{fig-hdisksize} we present a 
histogram showing the radial extent in kiloparsecs of the nuclear disks, which 
have a mean value of 0.24$\pm$0.31\,kpc and never extend beyond 1.1\,kpc. 
Therefore the bright stellar disks in the inactive galaxies are nuclear 
structures, similar or somewhat smaller in average spatial extent to the 
dust structures observed in the active galaxies. 

\section{Discussion\label{sec-discussion}}

In the previous section we identified two significant differences between 
active and inactive early-type galaxies. First, all active early-type 
galaxies have circumnuclear dust, while dust is only present in 26\% of 
inactive early-type galaxies. Second, 38\% of the inactive early-type 
galaxies have nuclear stellar disks, while they are detectable in almost 
none of the active, early-type galaxies. In the subsections below we 
discuss these results in the context of AGN fueling, the origin of the 
dust and stellar disks, and present a simple evolutionary scenario for 
the centers of early-type galaxies. 

\subsection{Ubiquitous Dust in Early-Type AGN Hosts} 

Dust is present in all of the active early-type galaxies, both in the matched 
and in the extended sample. In contrast, dust was only found in 27\% (7/26) 
of the inactive, matched sample and 26\% (9/34) of the extended, inactive 
sample. This strong correlation indicates that the presence of circumnuclear 
dust is a {\it requirement} for black hole accretion in early-type galaxies, 
while the presence of circumnuclear dust in some inactive galaxies indicates 
it is not a sufficient condition for activity. This strong correlation implies 
that the circumnuclear dust in these galaxies, most of which is hundreds of 
parsecs from the central black hole, is connected to or had a similar origin 
to the material currently fueling the active nucleus. 

As the circumnuclear dust observed in early-type, inactive galaxies almost 
always extends to the unresolved nucleus (tens of parsecs), this suggests that when 
dust is present in early-type galaxies, at least some dust reaches the nucleus 
relatively quickly compared to its destruction time. The ratio of inactive to
all early-type galaxies with circumnuclear dust in a representative (as
opposed to matched) sample multiplied by the dust lifetime could provide an 
estimate of how long it takes circumnuclear fuel to reach the nucleus. If we 
take the early-type galaxies in the Palomar survey as representative of 
all early-types, \citet{ho97b} found that approximately 50\% of all early-type
galaxies are AGN. Our matched sample of early-type active and inactive 
galaxies is therefore likely representative of all early-type galaxies and 
we infer that the time for some fuel to accrete onto the central black hole 
is approximately 20\% of the lifetime that dust is observable in the 
circumnuclear region. This percentage would be yet smaller if the dust in 
the circumnuclear region compromised the identification of some early-type 
galaxies as active. 

There are differing models for the origin and lifetime of dust in early-type 
galaxies. Unlike in late-type disks, there is not an obvious reservoir of dust 
on larger scales. The absence of such a dust reservoir indicates that the dust 
must be either created {\it in situ} or originate from outside of the 
galaxy. Stellar mass loss could produce the observed dust in early-type 
galaxies \citep{knapp92}. \citet{athey02} show that the mid-infrared dust 
emission from early-type galaxies is consistent with the expected stellar 
mass-loss rates, where the mass-loss rate is on order 1 \msun$yr^{-1}$ for a giant 
elliptical galaxy. The two difficulties with this process are that the dust 
should be produced in a uniform distribution that traces the stellar 
kinematics and the dust will gradually be destroyed via sputtering by the hot 
gas that pervades these galaxies. 
The dust distribution in a flattened disk is somewhat puzzling in the 
elliptical galaxies, such as NGC 4261, because they generally do not exhibit 
substantial rotation, although even initially slow rotation would lead 
to substantial rotation if the dust settled into the central kiloparsec from 
large scales. 
\citet{mathews03} model the evolution of 
the dusty gas originating in stellar winds and found cooling dust-rich gas 
could produce the observed circumnuclear dust, and in particular the observed 
clumping. However, even if dust cooling explains the observed clumpiness, 
stellar mass loss still can not explain why circumnuclear dust is only present 
in $\sim 50$\% of all early-type galaxies. While dust destruction via 
collisions with hot gas will occur in on order $10^8$ years 
\citep[e.g.,][]{mathews03}, mass loss will continually supply new dust to 
the interstellar medium. 
This difficulty with internal dust production is also reflected in the absence 
of a correlation between far-infrared and visible-wavelength luminosities in 
early-type galaxies, which also suggests the dust may have an external 
origin \citep{temi04}. 

Kinematic observations of neutral and ionized gas in early-type galaxies have 
led many authors to conclude that this material has an external origin because 
the gas kinematics differ from the stellar kinematics \citep[e.g.][and 
references therein]{bertola84,bertola92,sarzi06,morganti06}. The circumnuclear 
morphology of the dust also provides some additional constraints on the origin 
of the dust. In particular, the chaotic appearance suggests an external origin 
because internally-created dust might be too uniformly distributed, 
particularly in lenticulars, if the 
cooling time discussed above is not shorter than the dynamical time. However, 
the circumnuclear dust observed in late-type galaxies exhibits a similar range 
in morphologies and that material is generally not held to have an external 
origin. The morphology of the dust can also provide some constraints on 
the lifetime of the dust. \citet{lauer05} discuss how the different 
dust morphologies can be viewed as a "settling sequence," where dust that 
is not in a disk must settle toward the center of the galaxy in a 
dynamical time, or a few $10^7$ years \citep[see also][]{tran01,kleijn05}, 
while coherent, tightly wound dust disks must be at least a few dynamical 
times old. The age of such a dust disk would then be set by the dust 
destruction timescale mentioned above, or $\sim 10^8$ years. External dust 
input and a finite dust lifetime can plausibly explain why dust is not 
found in the centers of all early-type galaxies. 

More broadly, the relative number of galaxies with chaotic dust lanes, some 
nuclear dust spirals, and very coherent dust disks could provide constraints 
on the settling time of the dust if there is a natural progression toward 
more ordered dust as the dust settles toward the galaxy nucleus or is 
destroyed. If this model is correct, then the dust in inactive galaxies should 
be most chaotic or unsettled. Our data are consistent with this 
interpretation, with the exception of the tightly wound dust spiral in 
NGC~4526, although there are only a small number of tightly wound dust 
spirals in our full early-type galaxy sample. In fact, the relative scarcity 
of the tightly wound dust spirals presents a problem because the estimates 
of the dust settling time suggest that these structures should be longer-lived 
than the more commonly observed amorphous dust. As noted by \citet{lauer05}, 
this suggests the total dust lifetime must be close to the dynamical time 
of several times $10^7$ years and these structures can not be long lived. 

Both the internal and external dust models rely on a dust destruction 
timescale of at most $10^8$ years set by sputtering, while the observed 
fraction of dust and AGN in early-type galaxies indicates that about 50\% of 
early-type galaxies have dust at any time and this fraction approximately 
corresponds to the fraction with AGN. The main problem with a purely 
internal origin is that the dust creation from mass loss and the 
destruction from sputtering are continuous processes and all early-type 
galaxies should have some dust. The internal origin model therefore 
requires some more cataclysmic event to destroy the dust on approximately 
the same timescale required to reform the observed dust mass. In this 
case, the central AGN provides the most ready supply of the necessary energy to 
unbind or destroy the dust. The external origin, in contrast, requires 
continual dust replenishment via mergers with gas and dust-rich galaxies. 
Given the high fraction of early-type galaxies with dust, the frequency of 
these mergers must be comparable to the dust destruction time. In this case, 
the estimate of $10^8$ years appears to require an incredibly high merger 
rate with gas-rich dwarf galaxies. 
If this is the case, the dusty or active early-type galaxies may be 
preferentially in richer environments or have more companions. We checked 
this with measurements of the average local galaxy density and average 
projected angular separation to the nearest neighbor from \citep{ho97a} and 
did not find evidence for higher local density or closer companions in the 
active or dusty samples. 

\subsection{Nuclear Stellar Disks in Early-Type Galaxies} \label{sec:ndisks} 

Approximately a third of the inactive, early-type galaxies have structures 
that appear to be nuclear stellar disks. 
As noted in \S~\ref{sec-results}, these are likely to be stellar disks, 
rather than nuclear stellar bars, because they have the same position angle 
as their host galaxy. The true incidence of stellar disks in our sample 
is likely to be higher than observed and may be 100\% because most of these 
disks have very high axis ratios and we are most sensitive to stellar disks 
seen edge on. Many previous studies have found nuclear stellar disks in the 
central regions of elliptical and lenticular galaxies. \citet{ravindranath01} 
found such disks in 21\% of their 33 early-type galaxies, \citet{lauer05} 
found that nearly all of the power-law galaxies in their sample have stellar 
disks, and \citet{ferrarese06} find stellar disks in 13\% of a sample of 100 
early-type galaxies in the Virgo Cluster. Nuclear stellar disks are therefore 
a common structure, although only a few have been studied in great detail 
\citep{pizzella02,morelli04,kormendy05}.  

The origin of nuclear stellar disks has been a matter of recent debate, as 
these disks are not simple extensions of the large-scale disks to the centers  
of the galaxies. \citet{krajnovic04} argue that the nuclear disks could be the 
result of the infall of mass to the center of the galaxy driven by the secular 
evolution of a bar, galaxy mergers, or both. These authors also point out 
that none of the previously investigated nuclear stellar disk galaxies has an 
active nucleus, in agreement with our findings, although they do harbor a 
10$^{8-9}$\,M$_\sun$ SMBH, and argue that this makes them descendants of 
quasars that spent their fuel and turned off the central engine. Some known 
nuclear disks do exhibit strong evidence for an external origin, such as that 
in NGC~4698 \citep{pizzella02}, which is geometrically decoupled from the host 
galaxy. 

The ellipticity distribution of the five elliptical host galaxies with nuclear 
stellar disks may also have implications for the origin and detectability 
of these disks. As noted above, four of the five elliptical galaxies with 
nuclear stellar disks have very large axis ratios (three are classified 
E5 and one is classified `E6?'). Although this is a small sample, such large 
axis ratios are rare and suggests that nuclear stellar disks could 
preferentially exist in the subset of intrinsically flattened ellipticals 
and not just observed in the subset of the nuclear disks with a favorable 
viewing angle. As lower-luminosity ellipticals are more likely to be 
oblate or flattened \citep{anthony05} and tend to be faster rotators 
\citep{cappellari06}, the incidence of nuclear stellar disks should be 
reinvestigated as a function of elliptical galaxy properties, such as 
luminosity. 

\subsection{Evolution and timescales}

The nuclear stellar disks observed in early-type, inactive galaxies may share 
a common origin with the dusty disks observed in early-type AGN hosts. 
Once a nuclear stellar disk is formed in a galaxy, it provides a natural 
preferred plane of rotation for new cold gas and dust. New material, either 
infalling or formed {\it in situ} would likely settle onto the disk in several 
rotation periods. While the dust settles, it would likely have a relatively 
chaotic appearance at first, although eventually it would form a relatively 
regular disk of dust and may host star formation that would increase the mass 
of the stellar disk. \citet{kormendy05} observe cospatial nuclear stellar 
and dust disks in the dwarf elliptical NGC 4486A and suggest that these disks 
are toward the end stage of this evolutionary scenario, namely the observed 
stellar disk is forming from accreted gas and dust. 
This scenario is also supported by the recent study of \citet{ferrarese06}. 
These authors find evidence of recent star-formation 
associated with dusty disks but not with the irregular, unsettled dust lanes.  
They suggest a similar evolutionary scenario for dust settling in which dust is 
acquired externally, form large scale dust lanes that gradually evolve to 
dusty disks with associated star-formation, and then finally collapse to 
nuclear dusty disks associated with stellar disks. As star formation 
ceases and their colors become similar to the host galaxy, the disks 
then become more difficult to observe. 

Additional support for the evolutionary scenario above is the kinematic
evidence for inflow along nuclear spirals in at least one case: NGC\,1097.
Its nuclear dusty spiral has been revealed by near-IR observations
\citep{prieto05} and by structure maps in \citet{fathi06}. Using the
Integral-Field Unit of the Gemini Multi-Object Spectrograph, the latter
authors have found streaming motions along the nuclear spiral arms with inward
velocities of up to 50\,km\,s$^{-1}$ in the H$\alpha$ emitting gas throughout
the nuclear region. Another relevant finding on this galaxy is the young
obscured starburst discovered by \citet{sb05} very close to the nucleus
(within $\sim$10\,pc), in agreement with the suggestion that inflowing
gas and dust gives birth to stars in the nuclear dusty spiral or disk.
A key point in testing this evolutionary scenario is the evaluation of
the life cycles of the dusty and stellar disks. In the case of NGC\,1097,
the velocity observed for the streaming motions along the nuclear spiral allows
an estimate of a few Myr for the gas to flow from a few hundred parsecs to the
nucleus. This is also consistent with the above estimates that the dust may 
survive for on order several dynamical timescales. 

If substantial star formation does add to stellar nuclear disks each activity 
cycle, then there should be some stars associated with the disks younger than 
the characteristic or episodic timescale of the dust replenishment and AGN 
lifetime, or younger than $10^8$ years. While we do not have color information 
for our sample, a number of stellar disks are indeed found to be blue 
\citep{krajnovic04}, although many authors also argue that the colors are 
the same as those of the bulge \citep{erwin02}. Relatively old ages -- from 6 
to 15 Gyr -- have been reported from studies using broad-band colors and 
spectral indices \citep{krajnovic04,morelli04}. Nevertheless, a close look 
at these studies reveals that the methods used to date the disk stellar 
population are not very sensitive to the presence of a small young component 
in the middle of a luminous old bulge. In particular, the spectral indices are 
measured in a restricted interval (5000 $< \lambda <$ 5400\AA) and this 
wavelength range is not very sensitive to a small population of young stars and 
a blue spectral index should be used instead. 
The same applies for the broad-band colors, which usually do not include 
images in the blue. Estimates of the total cold gas mass also suggest that the 
number of new stars formed from the observed dust is likely to be small 
compared to the mass of the stellar disks. Detailed stellar population 
studies of nuclear stellar disks, combined with measurements of the star 
formation rates in the centers of dusty early-type galaxies, could test 
the consistency of this simple evolutionary model.

\section{Conclusions\label{sec-conclusions}}

We have used archival \hst\ WFPC2 images of 34 active and 34 inactive, 
early-type galaxies ($T<0$) to investigate if activity is correlated with the 
presence of circumnuclear dust. We found a strong correlation between the 
presence of circumnuclear (hundreds of parsecs scales) dust and nuclear 
activity: all early-type AGN hosts have circumnuclear dust, while only 26\% 
(nine of 34) early-type, inactive hosts have circumnuclear dust. 
This indicates that the presence of circumnuclear dust is a requirement, 
although not a sufficient condition, of black hole accretion in early-type 
galaxies. The circumnuclear dust in all of these early-type galaxies is 
typically confined to the central kiloparsec and is not observed on larger 
scales. The morphology of the circumnuclear dust is also typically complex, 
although a minority of galaxies ($<25$\%) show well-defined dusty disks with 
tightly wound spiral structure. Of the inactive galaxies, 38\% (13) have 
structures that appear to be nuclear stellar disks. As these structures appear 
to share the same position angle as the host galaxy, these structures are more 
likely to be stellar disks than nuclear stellar bars. We also investigated a 
sample of 31 active and 31 inactive late-type galaxies ($T \geq 0)$ and found 
that all of the galaxies show evidence for circumnuclear dust, independent of 
the presence of nuclear accretion. 

While our data demonstrate a clear connection between dust and AGN in 
early-type galaxies, the origin of the dust and therefore the nature of 
the AGN fueling mechanism remains unclear. We considered both internal 
and external origins for the circumnuclear dust, but both present 
significant challenges. Internal dust creation via stellar mass loss 
appears to be inconsistent with the absence of dust in approximately 50\% 
of all early-type galaxies. An external origin, in contrast, appears to 
require frequent mergers of small gas and dust-rich galaxies. The 
timescale of these mergers is set by the settling and destruction times 
of the dust, which may be as short as a few dynamical times for these 
kiloparsec or smaller structures, or $10^8$ years. 
The large observed fraction of early-type galaxies with nuclear stellar disks, 
the even larger fraction of early-type galaxies with dusty disks, and the 
relatively short lifetime of the dust, appears to require continual 
growth and replenishment of nuclear stellar disks. Our observations 
demonstrate that this process is also intimately linked with some accretion 
onto the central, supermassive black hole.  
Observations of the dust kinematics and mass, along with improved models of 
the dust settling and destruction at the centers of early-type galaxies, 
could resolve the origin of the dust and nuclear stellar disks, and reveal the 
fueling mechanism for these low-luminosity AGN.

\acknowledgements

We would like to thank Barbara Ryden for helpful discussions. 
RSL, TSB, MFS acknowledge support from the Brazilian institutions CNPq, 
CAPES and FAPERGS. Support for this work was also provided by NASA through 
grant AR-10677 from the Space Telescope Science Institute, which is operated 
by the Association of Universities for Research in Astronomy, Inc., under 
NASA contract NAS 5-26555.

\clearpage

\begin{deluxetable}{rccrccccccrrl}
\tabletypesize{\scriptsize}
\tablecaption{\hst\ data and general properties of the active sample \label{tab-active}}
\tablewidth{0pt}
\tablehead{

\colhead{(1)} & \colhead{(2)} & \colhead{(3)} & \colhead{(4)} & \colhead{(5)} & \colhead{(6)} & \colhead{(7)} & \colhead{(8)} & \colhead{(9)} & \colhead{(10)} & \colhead{(11)} & \colhead{(12)} & \colhead{(13)} \\
\colhead{NGC} & \colhead{WFPC2} & \colhead{Filter} & \colhead{Exposure} & \colhead{T} & \colhead{D} & \colhead{M$^{0}_{B_T}$} & \colhead{i} & \colhead{Activity} & \colhead{Pair} & \colhead{Dust} & \colhead{Disk} &\colhead{Comments} \\
\colhead{Number} & \colhead{Camera} & \colhead{} & \colhead{(sec)} & \colhead{} & \colhead{(Mpc)} & \colhead{(mag)} & \colhead{(deg)} & \colhead{Type} & \colhead{Number} & \colhead{(kpc)} & \colhead{(kpc)} & \colhead{} }

\tablecolumns{13}
\startdata

\sidehead{EARLY-TYPE MATCHED SAMPLE}
0315 & PC & F555W & 230 & -4 & 65.8 & -22.22 & 52 & L1.9 & 7619 & 0.41 & \nodata & d,i \\
2655 & PC & F547M & 300 & 0 & 24.4 & -21.12 & 34 & S2 & 4382 & $\times$ & \nodata & i,l \\
2768 & PC & F555W & 233 & -5 & 23.7 & -21.17 & \nodata & L2 & 4406 & 0.64 & \nodata & i,s \\
2787 & PC & F555W & 500 & -1 & 13 & -18.96 & 51 & L1.9 & 3384 & 0.61 & \nodata & d,s \\
2911 & WF & F547M & 460 & -2 & 42.2 & -20.88 & 40 & L2 & 5308 & 0.61 & \nodata & i,l \\
3226 & PC & F547M & 460 & -5 & 23.4 & -19.4 & \nodata & L1.9 & 5576 & 0.36 & \nodata & d \\
3414 & PC & F555W & 160 & -2 & 24.9 & -20.12 & 44 & L2 & 4371 & 0.15 & \nodata & s \\
3516 & PC & F555W & 1000 & -2 & 38.9 & -20.81 & 40 & S1.2 & 2950 & 0.9 & \nodata & i,s \\
3607 & PC & F547M & 260 & -2 & 19.9 & -20.7 & 62 & L2 & 4526 & 1.23 & \nodata & b?,d,s \\
3945 & PC & F555W & 600 & -1 & 22.5 & -20.41 & 50 & L2 & 4621 & $\times$ & \nodata & i \\
3998 & PC & F547M & 240 & -2 & 21.6 & -20.18 & 34 & L1.9 & 2300 & 0.43 & \nodata & i,l \\
4036 & PC & F547M & 300 & -3 & 24.6 & -20.46 & 69 & L1.9 & 4026 & $\times$ & \nodata & i,l \\
4111 & PC & F547M & 300 & -1 & 17 & -19.55 & 85 & L2 & 1023 & 0.25 & 0.57 & r \\
4138 & PC & F547M & 360 & -1 & 17 & -19.05 & 50 & S1.9 & 4694 & 0.21 & \nodata & l,s \\
4143 & PC & F606W & 280 & -2 & 17 & -19.25 & 52 & L1.9 & 4612 & 0.47 & \nodata & s \\
4261 & PC & F547M & 800 & -5 & 35.1 & -21.37 & \nodata & L2 & 3640 & 0.15 & \nodata & d \\
4278 & PC & F555W & 1000 & -5 & 9.7 & -18.96 & 22.53 & L1.9 & 3377 & $\times$ & \nodata & i \\
4293 & WF & F606W & 160 & 0 & 17 & -20.23 & 65 & L2 & 4469 & $\times$ & \nodata & l \\
4374 & PC & F547M & 1200 & -5 & 16.8 & -21.12 & \nodata & L2 & 4649 & 0.52 & \nodata & l \\
4486 & PC & F606W & 400 & -4 & 16.8 & -21.64 & \nodata & L2 & 4365 & $\times$ & \nodata & i \\
4550 & PC & F555W & 400 & -1.5 & 16.8 & -18.77 & 78 & L2 & 4570 & 0.77 & \nodata & i,s \\
4589 & PC & F555W & 1000 & -5 & 30 & -20.71 & \nodata & L2 & 5557 & $\times$ & \nodata & i,l \\
4636 & PC & F547M & 500 & -5 & 17 & -20.72 & \nodata & L1.9 & 0821 & $\times$ & \nodata & i \\
5077 & PC & F547M & 260 & -5 & 40.6 & -20.83 & \nodata & L1.9 & 4291 & $\times$ & \nodata & w \\
5273 & PC & F606W & 560 & -2 & 21.3 & -19.26 & 24 & S1.5 & 4379 & 0.14 & \nodata & l,s \\
7743 & PC & F606W & 560 & -1 & 24.4 & -19.78 & 32 & S2 & 4578 & 0.73 & \nodata & s \\
\sidehead{EARLY-TYPE EXTENDED SAMPLE}
0404 & PC & F606W & 280 & -3 & 2.4 & -15.98 & \nodata & L2 & 0221 & 0.04 & \nodata & s \\
1358 & PC & F606W & 500 & 0 & 53.6 & -20.95 & 38 & S2 & 7457 & 0.76 & \nodata & s \\
3166 & PC & F547M & 300 & 0 & 22 & -20.7 & 62 & L2 & 3115 & 0.92 & \nodata & i,l,s \\
3884 & PC & F547M & 460 & 0 & 91.6 & -21.55 & 51 & L1.9 & 4405 & 0.77 & \nodata & s \\
4203 & PC & F555W & 160 & -3 & 9.7 & -18.32 & 21 & L1.9 & 4245 & 0.23 & \nodata & s,l \\
4477 & PC & F606W & 160 & -2 & 16.8 & -19.83 & 24 & S2 & 3610 & 0.24 & \nodata & s \\
5548 & PC & F606W & 500 & 0 & 67 & -21.32 & 27 & S1.5 & 0507 & 2.54 & \nodata & s \\
6340 & PC & F606W & 300 & 0 & 22 & -20.04 & 24 & L2 & 3412 & 0.03 & \nodata & d \\
\sidehead{LATE-TYPE MATCHED SAMPLE}
1058 & PC & F606W & 160 & 5 & 9.1 & -18.25 & 21 & S2 & 3344 & $\times$ & \nodata & \nodata \\
1068 & PC & F606W & 560 & 3 & 14.4 & -21.32 & 32 & S1.8 & 3310 & $\times$ & \nodata & \nodata \\
1667 & PC & F606W & 500 & 5 & 61.2 & -21.52 & 40 & S2 & 2339 & $\times$ & \nodata & \nodata \\
2273 & PC & F606W & 500 & 0.5 & 28.4 & -20.25 & 41 & S2 & 2782 & $\times$ & \nodata & \nodata \\
2639 & PC & F606W & 500 & 1 & 42.6 & -20.96 & 54 & S1.9 & 0972 & $\times$ & \nodata & \nodata \\
3079 & PC & F606W & 560 & 7 & 20.4 & -21.14 & 90 & S2 & 3556 & $\times$ & \nodata & \nodata \\
3368 & PC & F606W & 280 & 2 & 8.1 & -19.74 & 47 & L2 & 3351 & $\times$ & \nodata & \nodata \\
3486 & PC & F606W & 560 & 5 & 7.4 & -18.58 & 43 & S2 & 3423 & $\times$ & \nodata & \nodata \\
3507 & PC & F606W & 80 & 3 & 19.8 & -19.86 & 32 & L2 & 4041 & $\times$ & \nodata & \nodata \\
3642 & PC & F547M & 300 & 4 & 27.5 & -20.74 & 34 & L1.9 & 5383 & $\times$ & \nodata & \nodata \\
3718 & PC & F547M & 300 & 1 & 17 & -19.96 & 62 & L1.9 & 4274 & $\times$ & \nodata & \nodata \\
3982 & PC & F606W & 500 & 3 & 17 & -19.47 & 30 & S1.9 & 4800 & $\times$ & \nodata & \nodata \\
4051 & PC & F606W & 500 & 4 & 17 & -20.41 & 43 & S1.2 & 6412 & $\times$ & \nodata & \nodata \\
4258 & PC & F606W & 560 & 4 & 6.8 & -20.63 & 70 & S1.9 & 2903 & $\times$ & \nodata & \nodata \\
4388 & PC & F606W & 560 & 3 & 16.8 & -20.34 & 83 & S1.9 & 4845 & $\times$ & \nodata & \nodata \\
4394 & WF & F606W & 320 & 3 & 16.8 & -19.62 & 27 & L2 & 0278 & $\times$ & \nodata & \nodata \\
4450 & PC & F555W & 520 & 2 & 16.8 & -20.38 & 43 & L1.9 & 4102 & $\times$ & \nodata & \nodata \\
4501 & PC & F606W & 600 & 3 & 16.8 & -21.27 & 59 & S2 & 2748 & $\times$ & \nodata & \nodata \\
4579 & PC & F547M & 726 & 3 & 16.8 & -20.84 & 38 & S1.9/L1.9 & 6217 & $\times$ & \nodata & \nodata \\
4639 & PC & F547M & 460 & 4 & 16.8 & -19.28 & 48 & S1.0 & 0864 & $\times$ & \nodata & \nodata \\
4651 & PC & F555W & 660 & 5 & 16.8 & -20.09 & 50 & L2 & 4567 & $\times$ & \nodata & \nodata \\
4698 & PC & F606W & 600 & 2 & 16.8 & -19.89 & 53 & S2 & 4380 & $\times$ & \nodata & \nodata \\
4772 & PC & F606W & 160 & 1 & 16.3 & -19.17 & 62 & L1.9 & 4448 & $\times$ & \nodata & \nodata \\
5033 & PC & F606W & 560 & 5 & 18.7 & -21.15 & 64 & S1.5 & 4559 & $\times$ & \nodata & \nodata \\
5377 & PC & F606W & 300 & 1 & 31 & -20.52 & 57 & L2 & 2146 & $\times$ & \nodata & \nodata \\
5566 & PC & F606W & 300 & 2 & 26.4 & -21.33 & 74 & L2 & 5806 & $\times$ & \nodata & \nodata \\
5985 & PC & F606W & 300 & 3 & 39.2 & -21.59 & 59 & L2 & 5248 & $\times$ & \nodata & \nodata \\
6500 & PC & F547M & 350 & 1.7 & 39.7 & -20.56 & 43 & L2 & 3504 & $\times$ & \nodata & \nodata \\
6951 & PC & F606W & 560 & 4 & 24.1 & -21.2 & 34 & S2 & 4254 & $\times$ & \nodata & \nodata \\
7217 & PC & F547M & 300 & 2 & 16 & -20.49 & 34 & L2 & 2775 & $\times$ & \nodata & \nodata \\
7479 & PC & F569W & 2000 & 5 & 32.4 & -21.33 & 41 & S1.9 & 4654 & $\times$ & \nodata & \nodata \\
\enddata
\tablecomments{(3) Pass-band \hst\ filter used in observation; (5--9) Galaxy T type, distance in Mpc, B-band absolute magnitude, inclination, and activity type from \citep{ho97a}, S indicates a Seyfert and L indicates a LINER; (10) NGC number of the control sample pair; (11) Main innermost dust structure semi-major axis size in kpc, a $\times$ sign means some dust structure is present but is difficult to be measured; (12) Stellar disk radial extent in kpc; (13) Comments: b = bar, d = dust disk, i = irregular dust, l = dust lanes or filaments, r = dust ring, s = dust spiral, w = weak feature, ? = uncertain classification.}
\end{deluxetable}

\clearpage

\begin{deluxetable}{rccrccccccrrl}
\tabletypesize{\scriptsize}
\tablecaption{\hst\ data and general properties of control sample \label{tab-control}}
\tablewidth{0pt}
\tablehead{

\colhead{(1)} & \colhead{(2)} & \colhead{(3)} & \colhead{(4)} & \colhead{(5)} & \colhead{(6)} & \colhead{(7)} & \colhead{(8)} & \colhead{(9)} & \colhead{(10)} & \colhead{(11)} & \colhead{(12)} & \colhead{(13)} \\
\colhead{NGC} & \colhead{WFPC2} & \colhead{Filter} & \colhead{Exposure} & \colhead{T} & \colhead{D} & \colhead{M$^{0}_{B_T}$} & \colhead{i} & \colhead{Activity} & \colhead{Pair} & \colhead{Dust} & \colhead{Disk} &\colhead{Comments} \\
\colhead{Number} & \colhead{Camera} & \colhead{} & \colhead{(sec)} & \colhead{} & \colhead{(Mpc)} & \colhead{(mag)} & \colhead{(deg)} & \colhead{Type} & \colhead{Number} & \colhead{(kpc)} & \colhead{(kpc)} & \colhead{} }

\tablecolumns{13}
\startdata

\sidehead{EARLY-TYPE MATCHED SAMPLE}
0821 & PC & F555W & 304 & -5 & 23.2 & -20.11 & \nodata & A & 4636 & \nodata & 0.2 & \nodata \\
1023 & PC & F555W & 420 & -3 & 10.5 & -20.03 & 73 & A & 4111 & \nodata & 0.069 & \nodata \\
2300 & PC & F555W & 304 & -2 & 31 & -20.69 & 44 & A & 3998 & \nodata & \nodata & \nodata \\
2950 & PC & F555W & 500 & -2 & 23.3 & -20.04 & 50 & A & 3516 & \nodata & \nodata & b \\
3377 & PC & F555W & 160 & -5 & 8.1 & -18.47 & \nodata & A & 4278 & 0.26 & 0.071 & l \\
3384 & PC & F555W & 380 & -3 & 8.1 & -18.79 & 65 & A & 2787 & \nodata & 0.034 & \nodata \\
3640 & PC & F555W & 400 & -5 & 24.2 & -20.73 & \nodata & A & 4261 & \nodata & \nodata & \nodata \\
4026 & PC & F555W & 400 & -2 & 17 & -19.56 & 81 & A & 4036 & $\times$ & 0.071 & i,l \\
4291 & PC & F555W & 304 & -5 & 29.4 & -20.09 & \nodata & A & 5077 & \nodata & \nodata & \nodata \\
4365 & PC & F555W & 1000 & -5 & 16.8 & -20.64 & \nodata & A & 4486 & \nodata & \nodata & \nodata \\
4371 & PC & F606W & 80 & -1 & 16.8 & -19.51 & 57 & A & 3414 & 0.05 & \nodata & d \\
4379 & PC & F555W & 160 & -2.5 & 16.8 & -18.6 & 32 & A & 5273 & \nodata & \nodata & \nodata \\
4382 & PC & F555W & 568 & -1 & 16.8 & -21.14 & 40 & A & 2655 & \nodata & \nodata & \nodata \\
4406 & PC & F555W & 500 & -5 & 16.8 & -21.39 & \nodata & A & 2768 & 0.02 & \nodata & w \\
4469 & PC & F606W & 160 & 0 & 16.8 & -19.31 & 73 & H & 4293 & 0.57 & \nodata & i,l \\
4526 & PC & F555W & 520 & 0 & 16.8 & -19.31 & 74 & H & 3607 & 1.21 & \nodata & s \\
4570 & PC & F555W & 200 & -2 & 16.8 & -19.33 & 76 & A & 4550 & \nodata & 0.119 & \nodata \\
4578 & PC & F606W & 80 & -2 & 16.8 & -18.96 & 43 & A & 7743 & \nodata & \nodata & \nodata \\
4612 & PC & F606W & 80 & -2 & 16.8 & -18.75 & 38 & A & 4143 & \nodata & \nodata & \nodata \\
4621 & PC & F555W & 330 & -5 & 16.8 & -20.6 & \nodata & A & 3945 & \nodata & 0.274 & \nodata \\
4649 & PC & F555W & 1050 & -5 & 16.8 & -21.43 & \nodata & A & 4374 & \nodata & \nodata & \nodata \\
4694 & PC & F606W & 500 & -2 & 16.8 & -19.08 & 63 & H & 4138 & $\times$ & \nodata & i \\
5308 & PC & F555W & 400 & -3 & 32.4 & -20.13 & \nodata & A & 2911 & \nodata & 0.993 & \nodata \\
5557 & PC & F555W & 500 & -5 & 42.6 & -21.17 & \nodata & A & 4589 & \nodata & \nodata & \nodata \\
5576 & PC & F555W & 1600 & -5 & 26.4 & -20.34 & \nodata & A & 3226 & \nodata & \nodata & \nodata \\
7619 & PC & F555W & 1100 & -5 & 50.7 & -21.6 & \nodata & A & 0315 & \nodata & 0.313 & \nodata \\
\sidehead{EARLY-TYPE EXTENDED SAMPLE}
0221 & PC & F555W & 26 & -6 & 0.7 & -15.51 & \nodata & A & 404 & \nodata & 0.004 & \nodata \\
0507 & PC & F555W & 567 & -2 & 65.7 & -21.96 & \nodata & A & 5548 & \nodata & \nodata & \nodata \\
3115 & PC & F555W & 280 & -3 & 6.7 & -19.39 & 73 & A & 1358 & \nodata & 0.067 & \nodata \\
3412 & PC & F606W & 80 & -2 & 8.1 & -18.2 & 57 & A & 3166 & \nodata & 0.063 & \nodata \\
3610 & PC & F555W & 2400 & -5 & 29.2 & -20.79 & \nodata & A & 6340 & \nodata & 0.895 & \nodata \\
4245 & PC & F606W & 160 & 0 & 9.7 & -17.92 & 41 & H & 4203 & 0.19 & \nodata & r \\
4405 & PC & F606W & 160 & 0 & 31.5 & -19.63 & 51 & H & 3884 & $\times$ & \nodata & \nodata \\
7457 & PC & F555W & 1050 & -3 & 12.3 & -18.69 & 59 & A & 4477 & \nodata & \nodata & \nodata \\
\sidehead{LATE-TYPE MATCHED SAMPLE}
0278 & WF & F606W & 320 & 3 & 11.8 & -19.69 & 18 & H & 4394 & $\times$ & \nodata & \nodata \\
0864 & PC & F606W & 560 & 5 & 20 & -20.25 & 41 & H & 4639 & $\times$ & \nodata & \nodata \\
0972 & PC & F606W & 600 & 2 & 21.4 & -20.17 & 61 & H & 2639 & $\times$ & \nodata & \nodata \\
2146 & PC & F606W & 560 & 2 & 17.2 & -20.6 & 57 & H & 5377 & $\times$ & \nodata & \nodata \\
2339 & PC & F606W & 600 & 4 & 30.9 & -20.97 & 41 & H & 1667 & $\times$ & \nodata & \nodata \\
2748 & PC & F606W & 600 & 4 & 23.8 & -20.29 & 70 & H & 4501 & $\times$ & \nodata & \nodata \\
2775 & PC & F606W & 300 & 2 & 17 & -20.34 & 40 & A & 7217 & $\times$ & \nodata & \nodata \\
2782 & PC & F555W & 460 & 1 & 37.3 & -20.85 & 43 & H & 2273 & $\times$ & \nodata & \nodata \\
2903 & PC & F606W & 560 & 4 & 6.3 & -19.89 & 63 & H & 4258 & $\times$ & \nodata & \nodata \\
3310 & PC & F606W & 500 & 4 & 18.7 & -20.41 & 40 & H & 1068 & $\times$ & \nodata & \nodata \\
3344 & PC & F606W & 160 & 4 & 6.1 & -18.43 & 24 & H & 1058 & $\times$ & \nodata & \nodata \\
3351 & PC & F606W & 560 & 3 & 8.1 & -19.28 & 48 & H & 3368 & $\times$ & \nodata & \nodata \\
3423 & PC & F606W & 160 & 6 & 10.9 & -18.81 & 32 & H & 3486 & $\times$ & \nodata & \nodata \\
3504 & PC & F606W & 500 & 2 & 26.5 & -20.61 & 40 & H & 6500 & $\times$ & \nodata & \nodata \\
3556 & PC & F606W & 160 & 6 & 14.1 & -20.92 & 80 & H & 3079 & $\times$ & \nodata & \nodata \\
4041 & WF & F606W & 320 & 4 & 22.7 & -20.02 & 21 & H & 3507 & $\times$ & \nodata & \nodata \\
4102 & PC & F606W & 600 & 3 & 17 & -19.54 & 56 & H & 4450 & $\times$ & \nodata & \nodata \\
4254 & PC & F606W & 560 & 5 & 16.8 & -21.03 & 30 & H & 6951 & $\times$ & \nodata & \nodata \\
4274 & PC & F555W & 296 & 2 & 9.7 & -19.35 & 71 & H & 3718 & $\times$ & \nodata & \nodata \\
4380 & PC & F606W & 560 & 3 & 16.8 & -19.06 & 58 & H & 4698 & $\times$ & \nodata & \nodata \\
4448 & PC & F606W & 160 & 2 & 9.7 & -18.65 & 71 & H & 4772 & $\times$ & \nodata & \nodata \\
4559 & PC & F606W & 160 & 6 & 9.7 & -20.17 & 68 & H & 2033 & $\times$ & \nodata & \nodata \\
4567 & PC & F606W & 160 & 4 & 16.8 & -19.34 & 48 & H & 4651 & $\times$ & \nodata & \nodata \\
4654 & PC & F606W & 160 & 6 & 16.8 & -20.38 & 56 & H & 7479 & $\times$ & \nodata & \nodata \\
4800 & PC & F606W & 160 & 3 & 15.2 & -18.78 & 43 & H & 3982 & $\times$ & \nodata & \nodata \\
4845 & PC & F606W & 160 & 2 & 15.6 & -19.55 & 79 & H & 4388 & $\times$ & \nodata & \nodata \\
5248 & PC & F547M & 900 & 4 & 22.7 & -21.15 & 44 & H & 5985 & $\times$ & \nodata & \nodata \\
5383 & PC & F606W & 560 & 3 & 37.8 & -20.94 & 32 & H & 3642 & $\times$ & \nodata & \nodata \\
5806 & PC & F606W & 600 & 3 & 28.5 & -20.43 & 61 & H & 5566 & $\times$ & \nodata & \nodata \\
6217 & PC & F606W & 500 & 4 & 23.9 & -20.23 & 34 & H & 4579 & $\times$ & \nodata & \nodata \\
6412 & PC & F606W & 560 & 5 & 23.5 & -19.78 & 30 & H & 4051 & $\times$ & \nodata & \nodata \\
\enddata
\tablecomments{(3) Pass-band \hst\ filter used in observation; (5--9) Galaxy T type, distance in Mpc, B-band absolute magnitude, inclination, and activity type from \citep{ho97a}, S indicates a Seyfert and L indicates a LINER; (10) NGC number of the control sample pair; (11) Main innermost dust structure semi-major axis size in kpc, a $\times$ sign means some dust structure is present but is difficult to be measured; (12) Stellar disk radial extent in kpc; (13) Comments: b = bar, d = dust disk, i = irregular dust, l = dust lanes or filaments, r = dust ring, s = dust spiral, w = weak feature, ? = uncertain classification.}
\end{deluxetable}

\clearpage

\begin{figure}
\plottwo{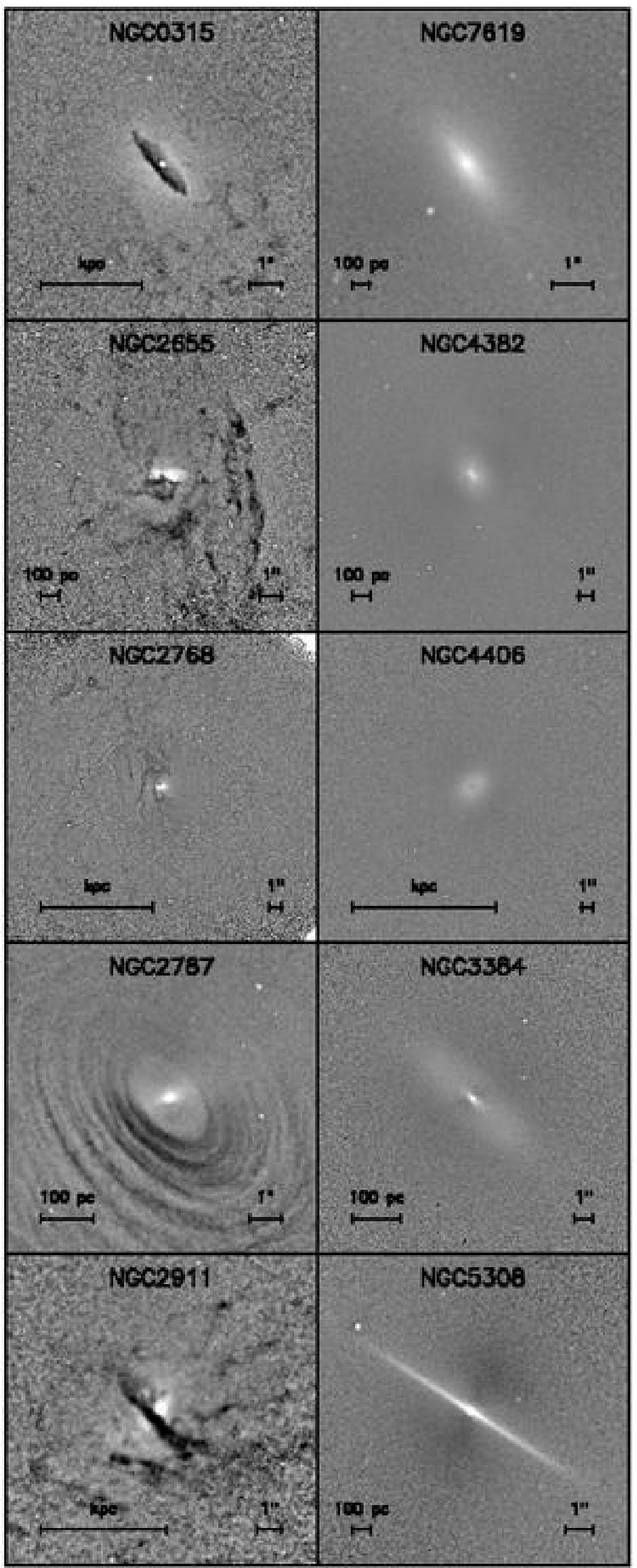}{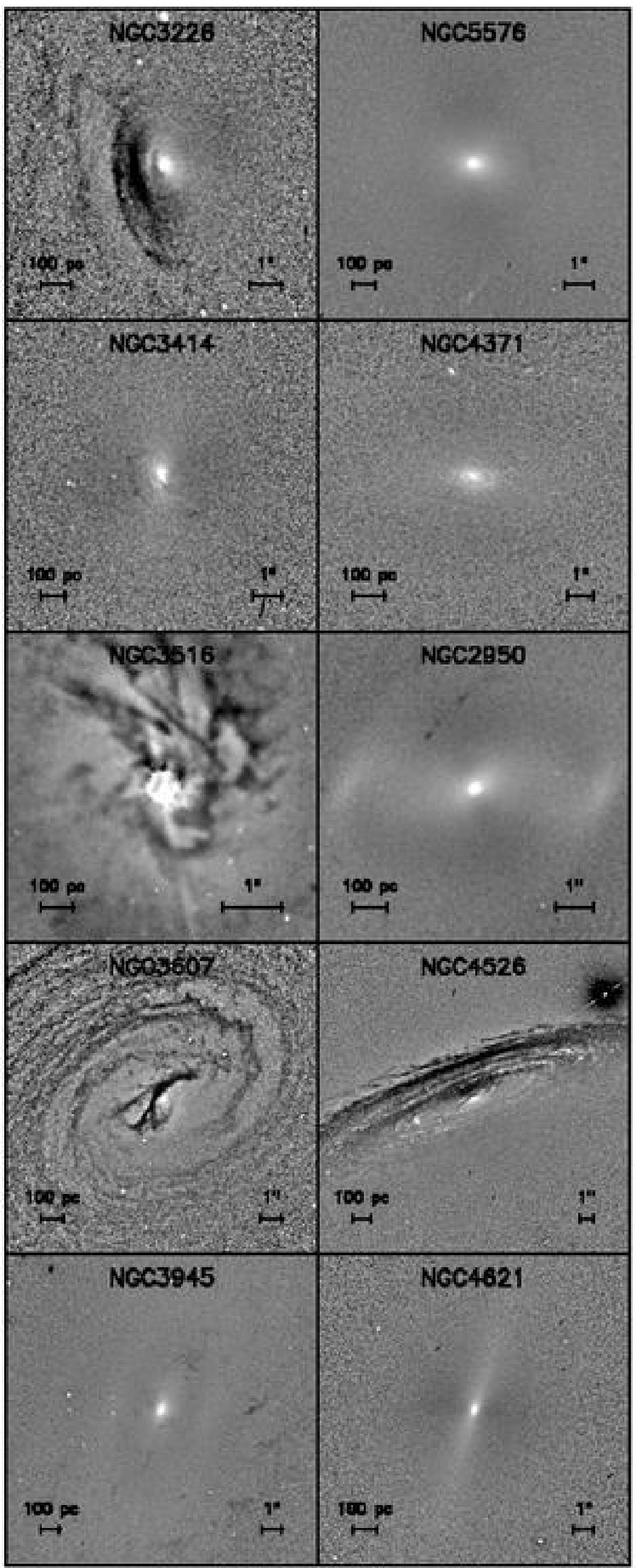}
\caption{Structure maps for the matched sample of early-type galaxies. Each image covers 5\% of the galaxy D25, and is rotated so that North is up and East is to the left. In each panel the left column displays the active galaxy and the right column the pair-matched control galaxy.\label{fig-stmap1}}
\end{figure} 

\setcounter{figure}{2}

\begin{figure}
\plottwo{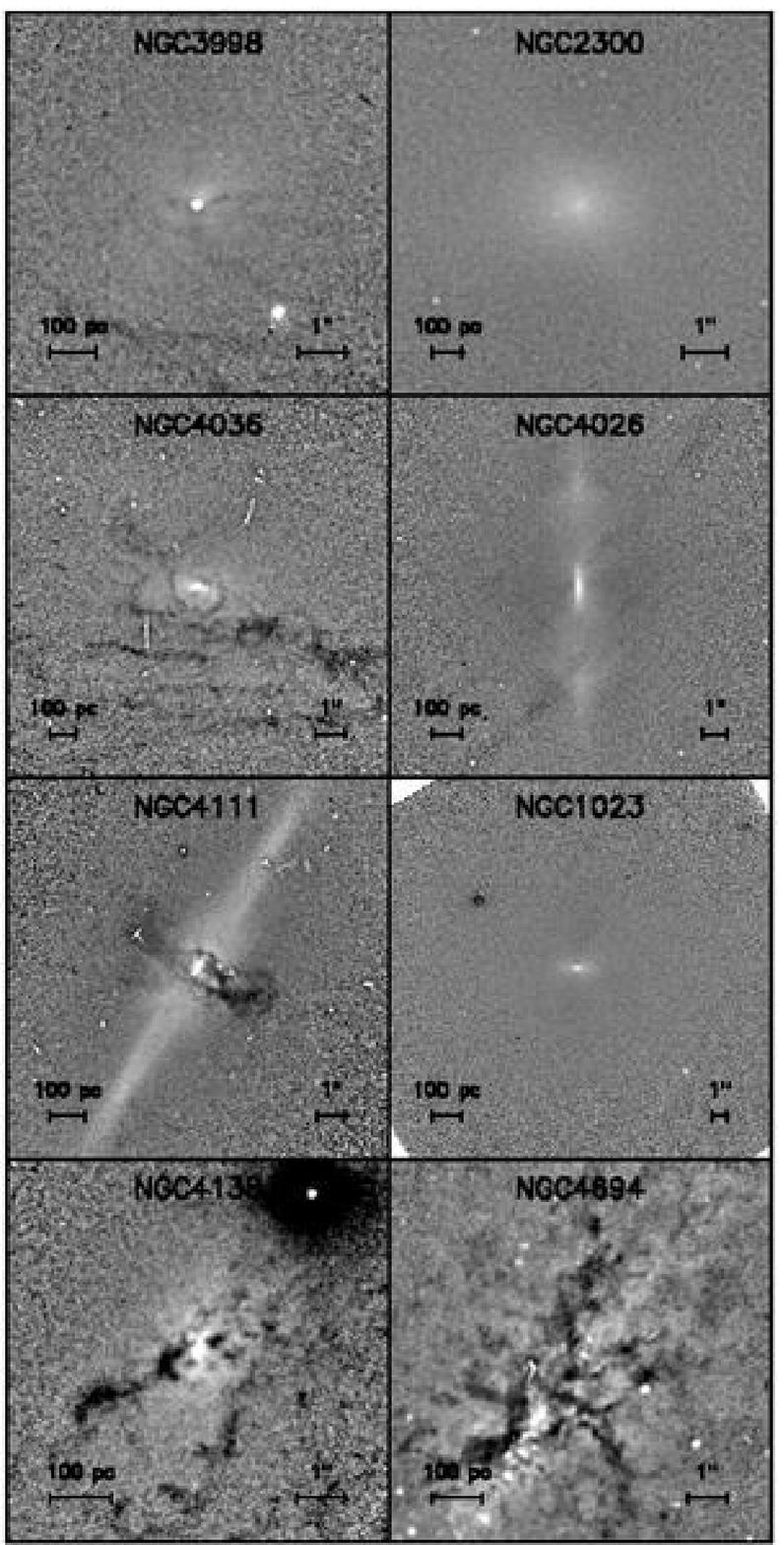}{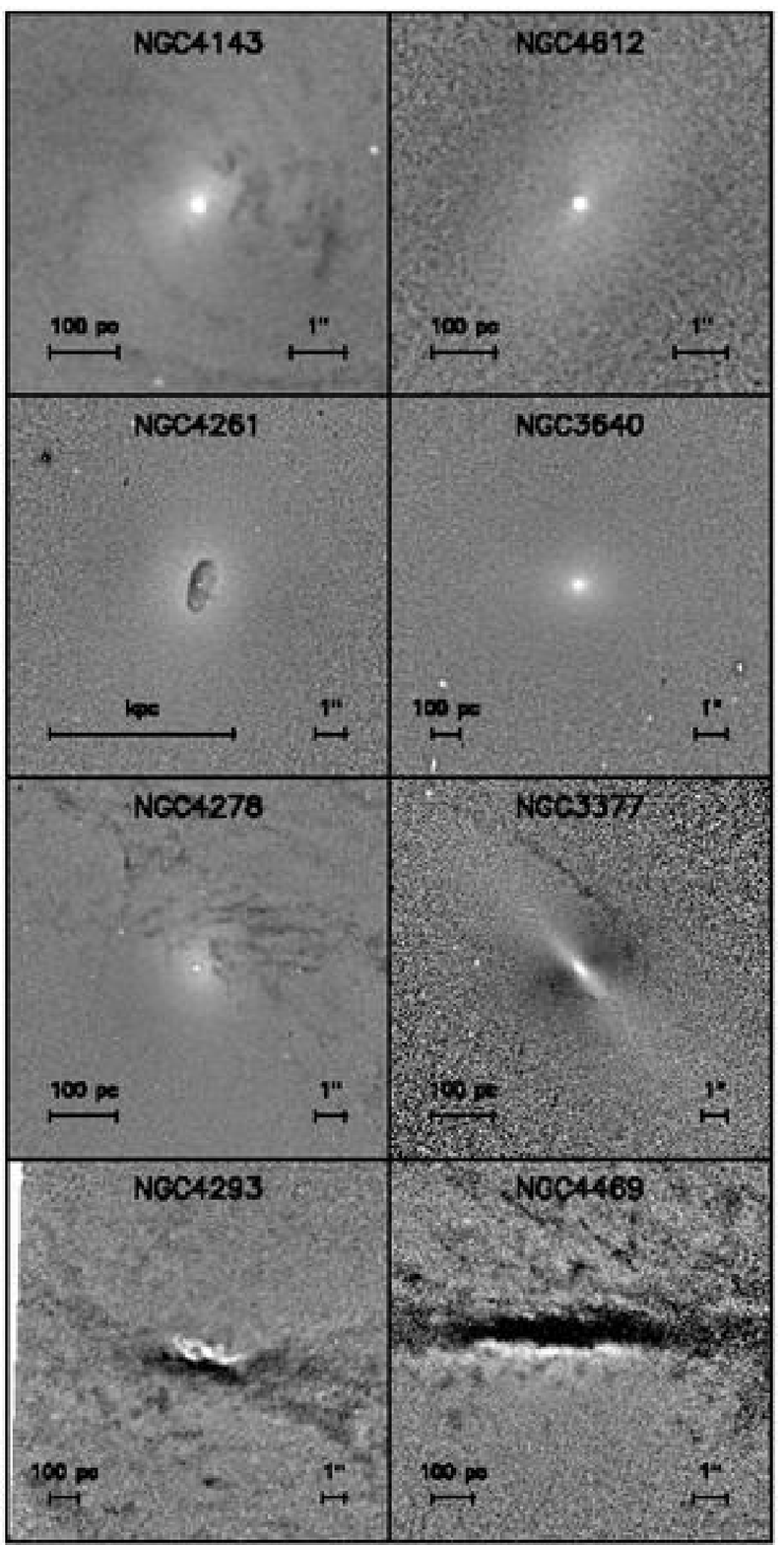}
\caption{continued}
\end{figure} 

\setcounter{figure}{2}

\begin{figure}
\plottwo{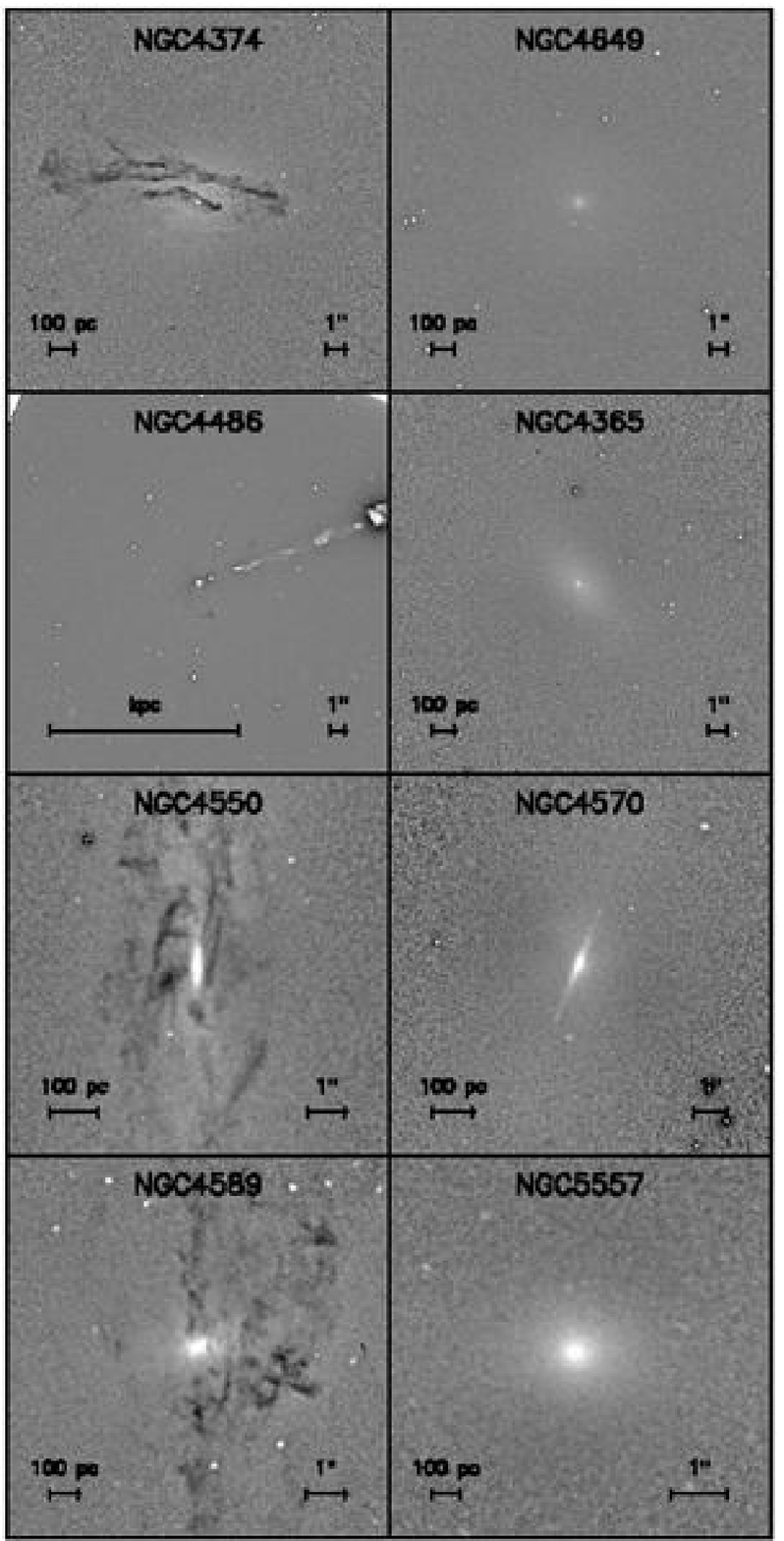}{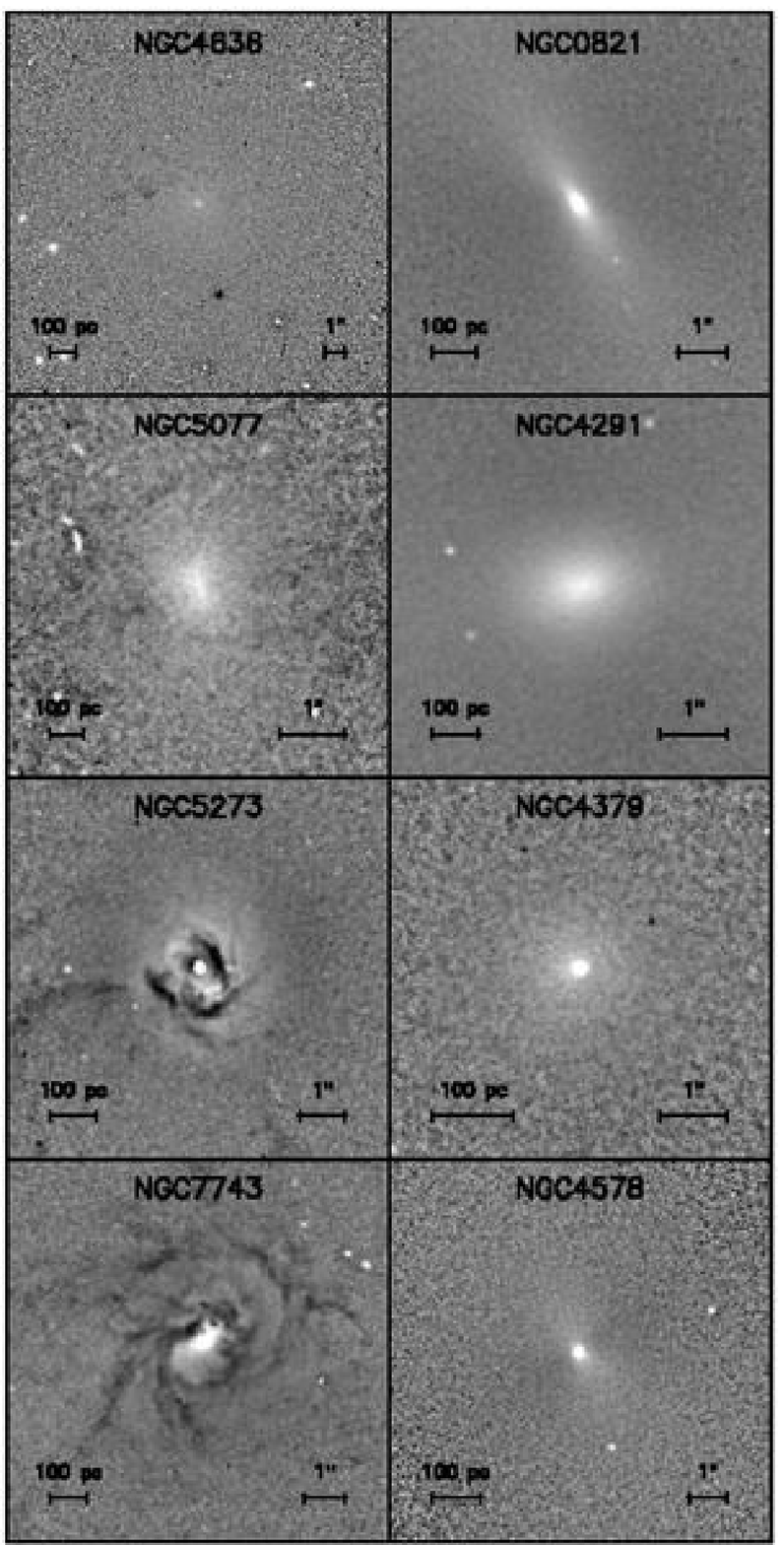}
\caption{continued\label{fig-stmap3}}
\end{figure} 

\begin{figure}
\plottwo{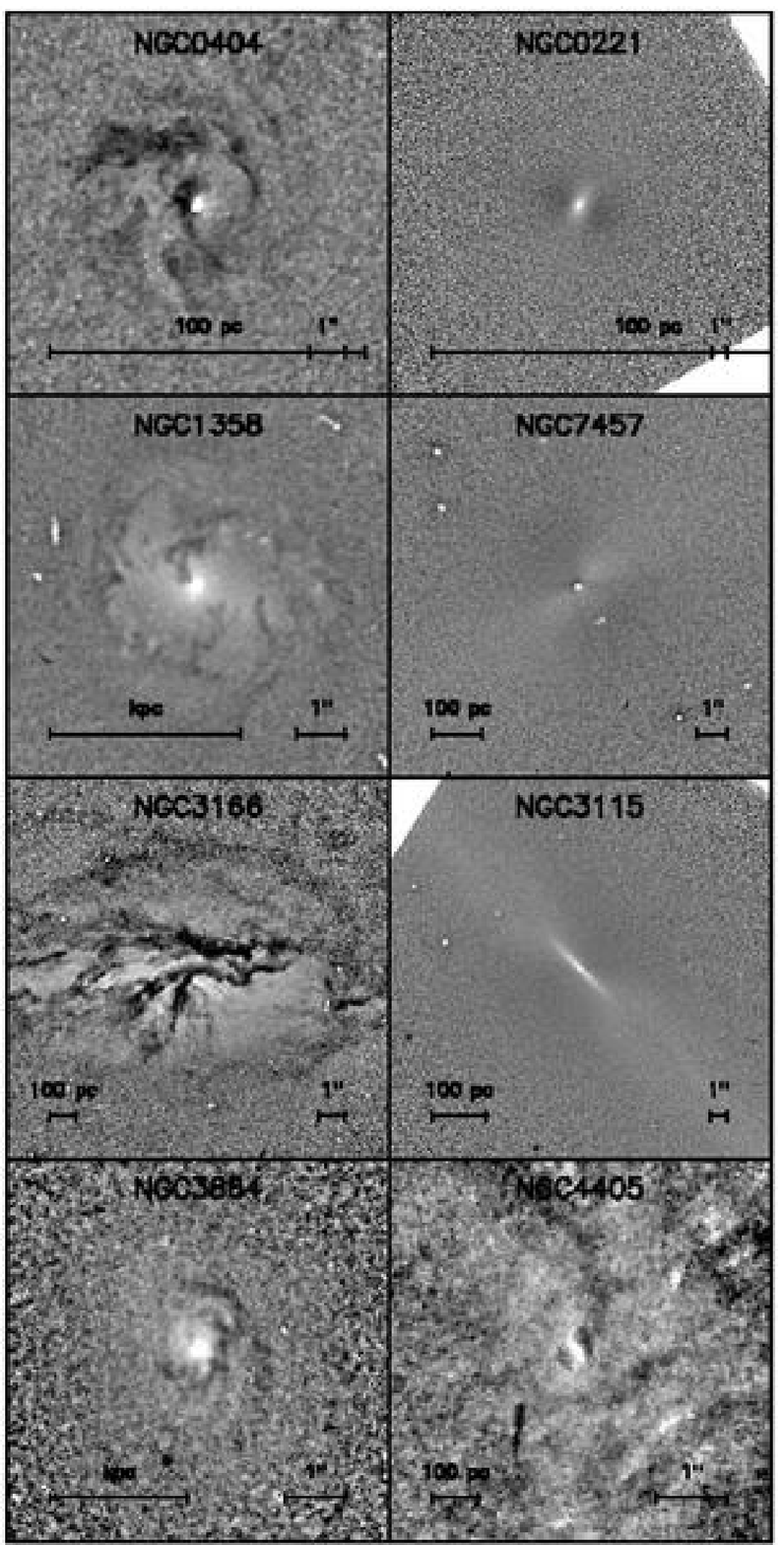}{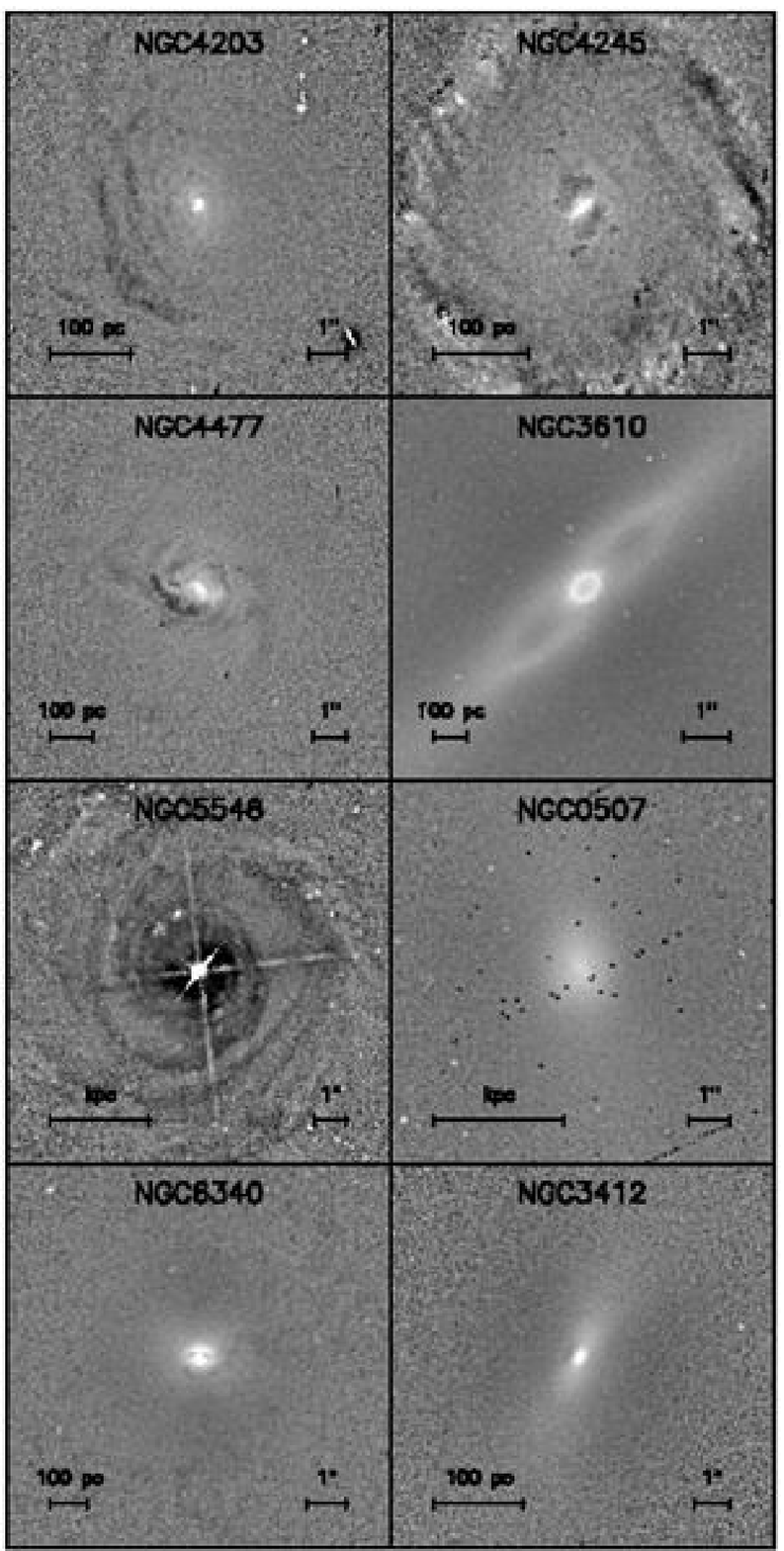}
\caption{Structure maps for the extended sample of early-type galaxies. Images are presented as in
Fig. \ref{fig-stmap1}.\label{fig-stmap4}}
\end{figure} 

\begin{figure}
\plottwo{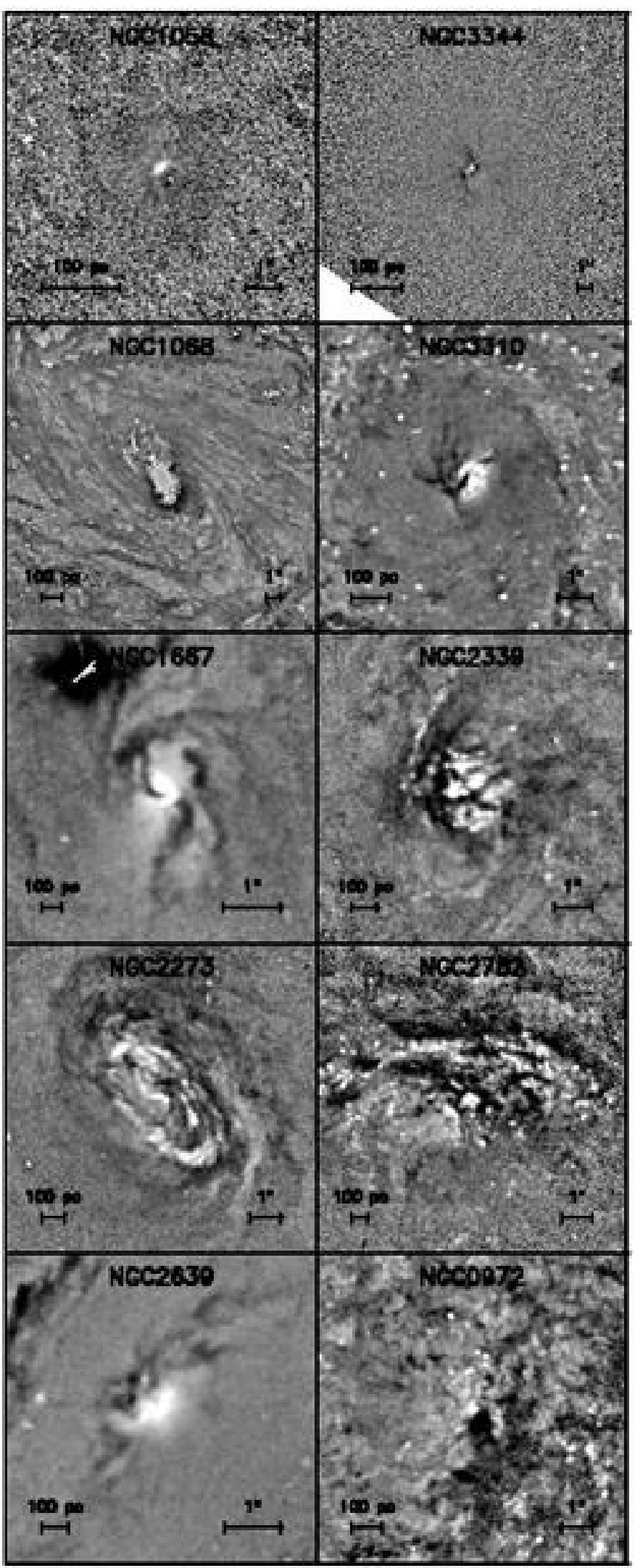}{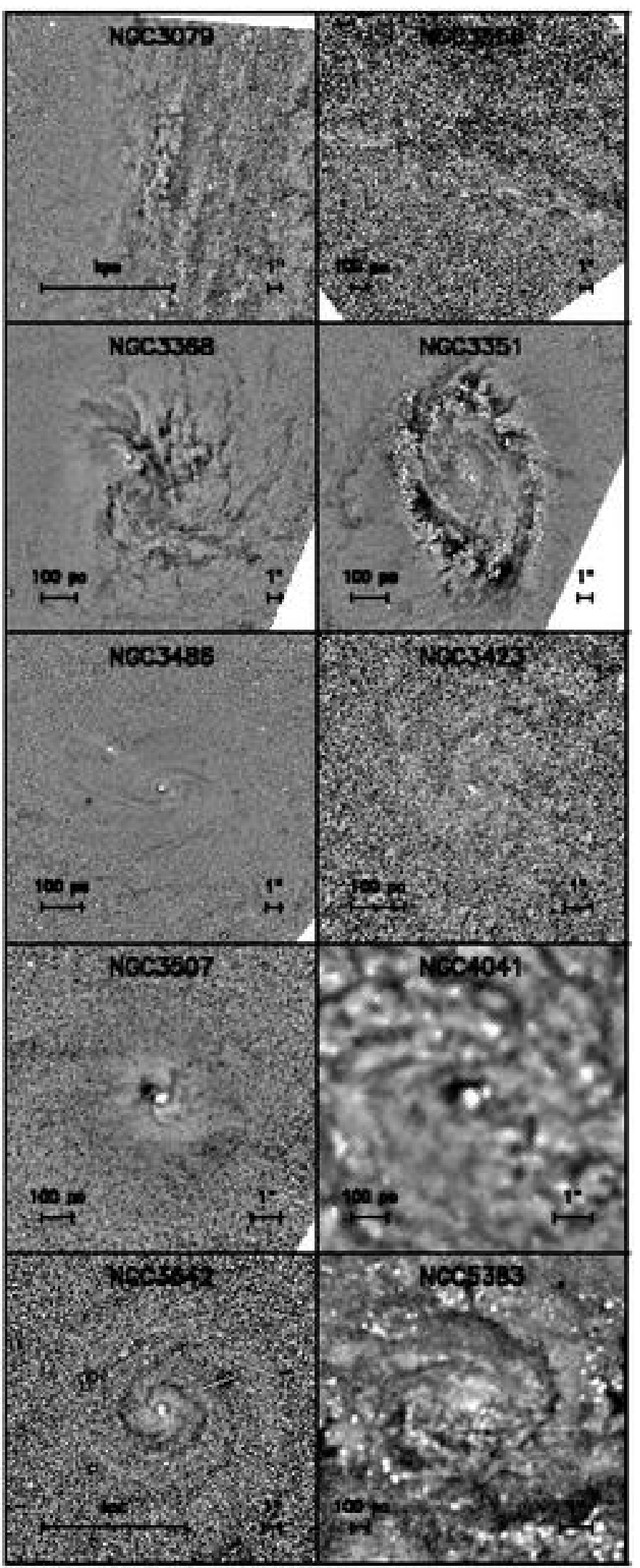}
\caption{Structure maps for the matched sample of late-type galaxies. Images are presented as in Fig. \ref{fig-stmap1}.\label{fig-stmap5}}
\end{figure}

\setcounter{figure}{4}

\begin{figure}
\plottwo{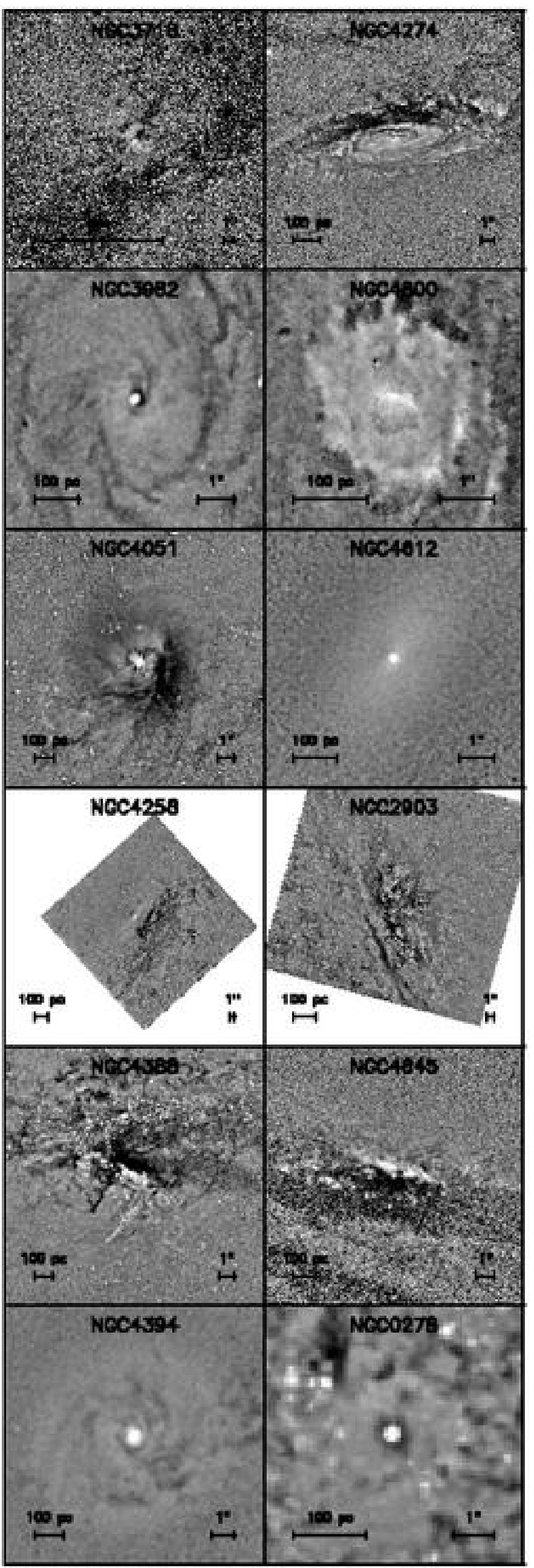}{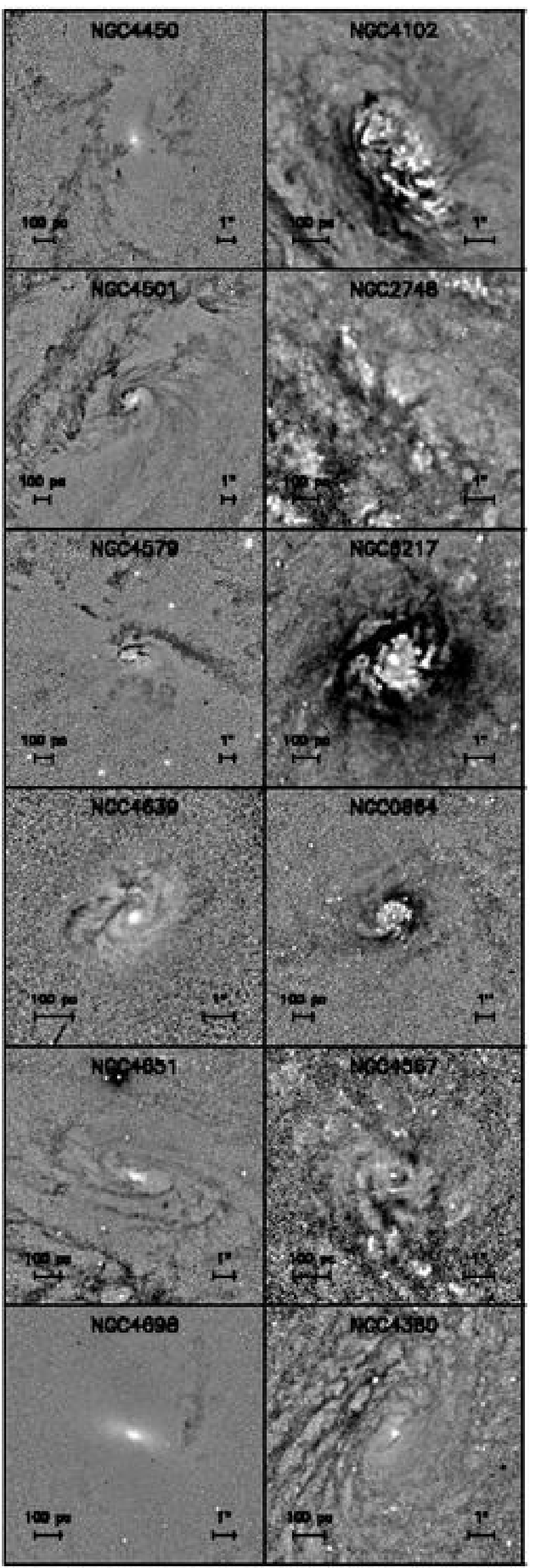}
\caption{continued}
\end{figure} 

\setcounter{figure}{4}

\begin{figure}
\plottwo{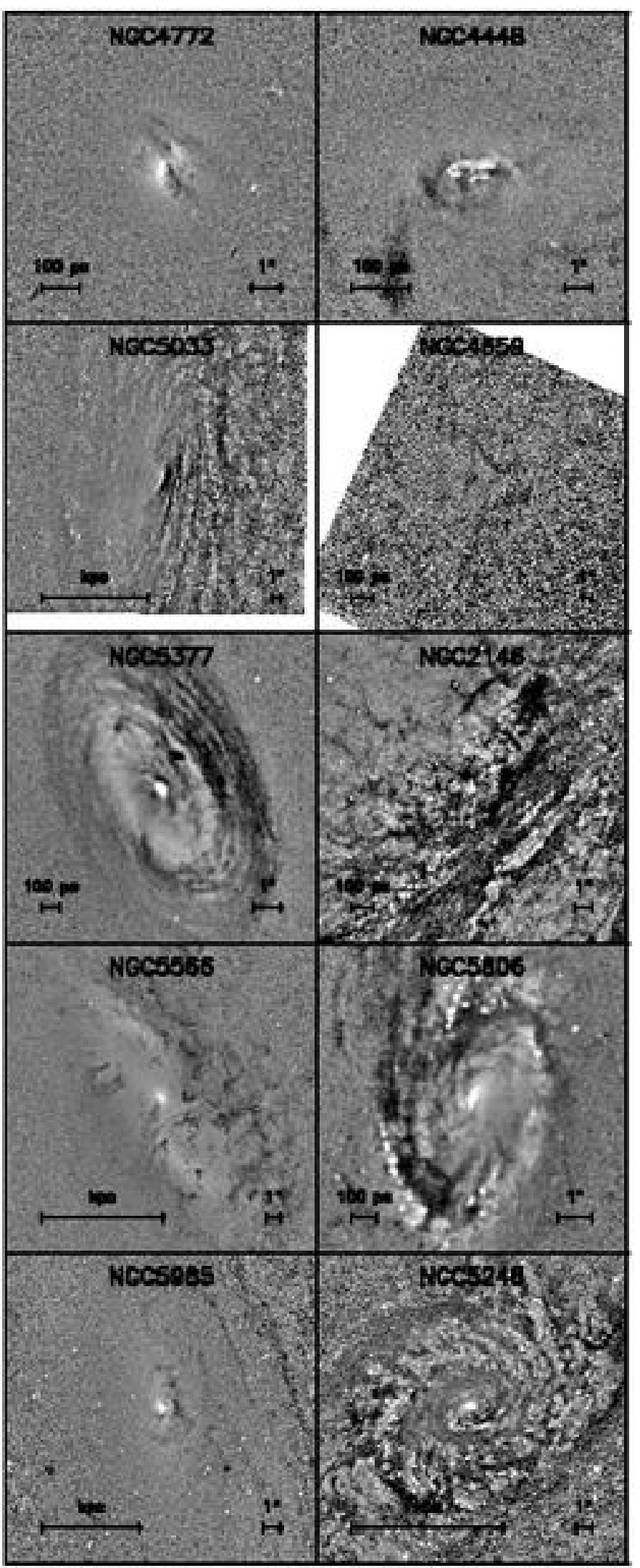}{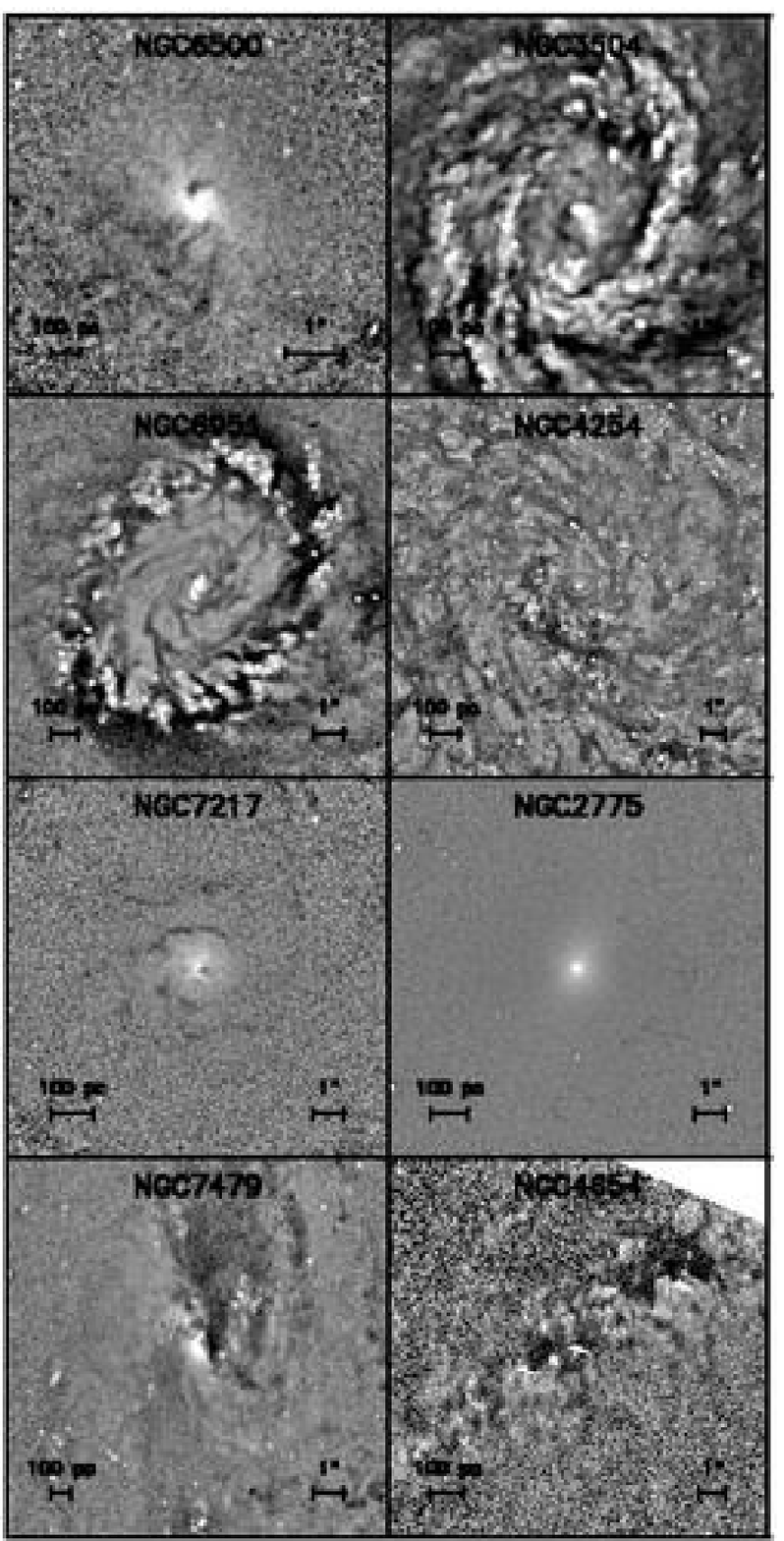}
\caption{continued\label{fig-stmap7}}
\end{figure} 

\begin{figure}
\epsscale{0.9}
\plotone{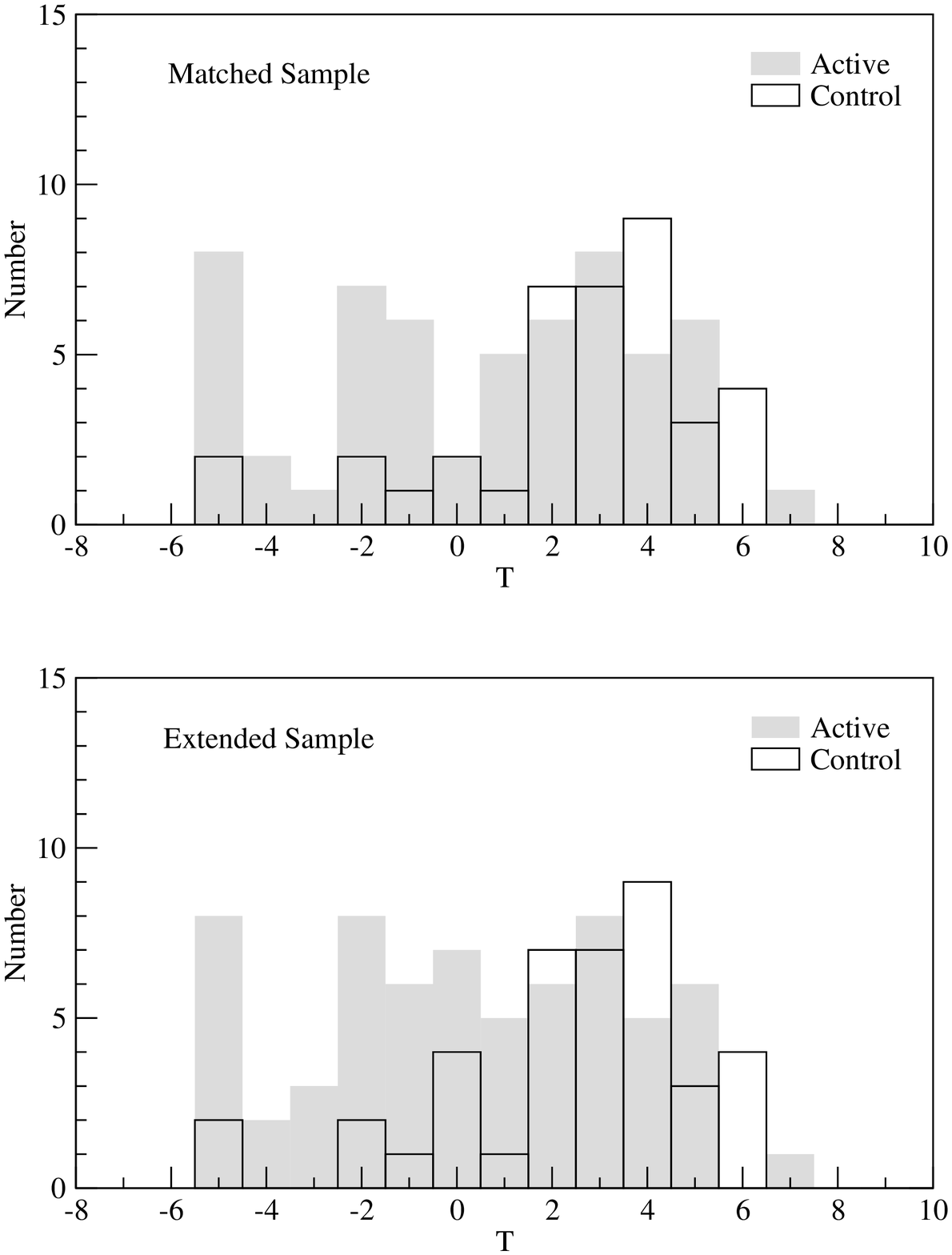}
\caption{Presence of dust structure in the matched (upper panel) and extended (lower panel) sample as a function of Hubble type for the active and control galaxies.\label{fig-hstructure}}
\end{figure}

\begin{figure}
\epsscale{0.9}
\plotone{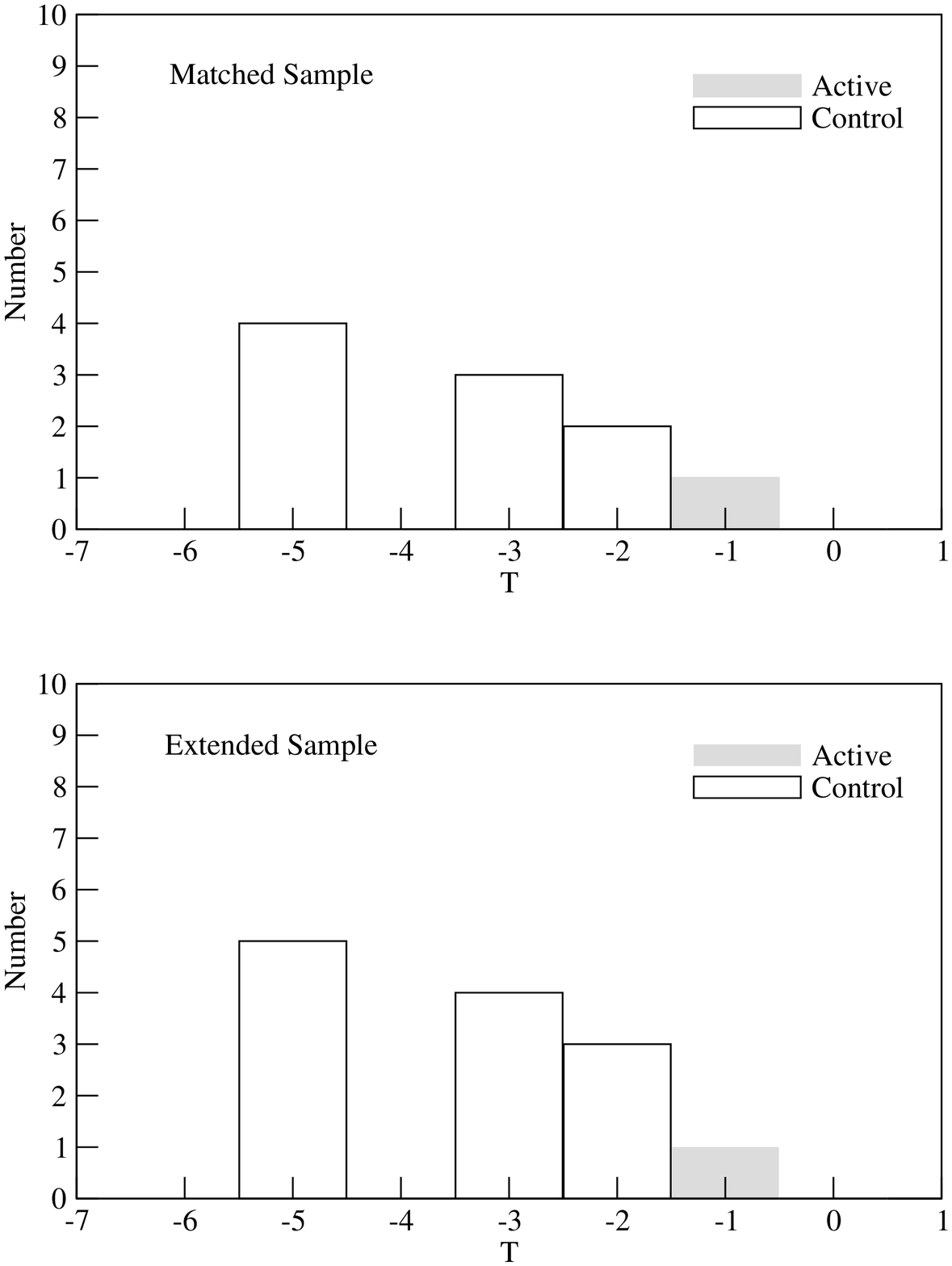}
\caption{Presence of stellar disks in the matched (upper panel) and extended (lower panel) sample as a function of Hubble type for the active and control galaxies.\label{fig-hdisk}}
\end{figure}

\begin{figure}
\epsscale{0.9}
\plotone{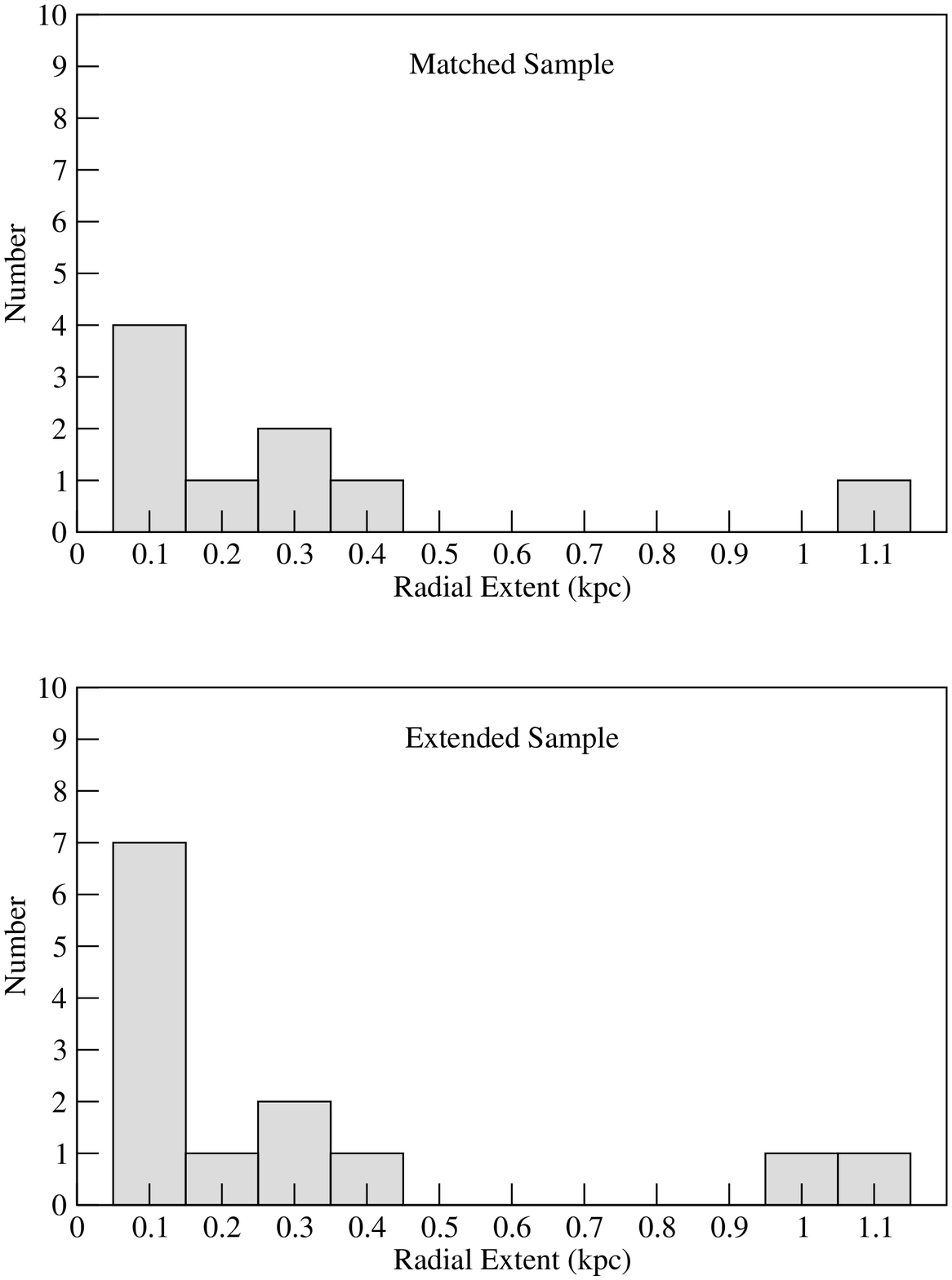}
\caption{Stellar disk's radial extent in kpc for the matched (upper panel) and extended (lower panel) sample.\label{fig-hdisksize}}
\end{figure}

\end{document}